\documentclass[pre,aps,twocolumn,showpacs,amsmath,amssymb,floatfix,superscriptaddress,citeautoscript]{revtex4-1}
\usepackage{graphicx}
\usepackage{dcolumn}
\usepackage{bm}
\usepackage{amsmath}
\usepackage{amssymb}
\usepackage{amssymb, verbatim}
\usepackage[normal]{subfigure}

\usepackage{hyperref}
\hypersetup{
        colorlinks=true,
}

\makeatletter

\usepackage{color}

\begin{document}

\title{Keldysh approach to periodically driven systems with a fermionic bath:
non-equilibrium steady state, proximity effect and dissipation}

\author{Dong E. Liu}
\affiliation{Station Q, Microsoft Research, Santa Barbara, California 93106-6105, USA}
\author{Alex Levchenko}
\affiliation{Department of Physics, University of Wisconsin-Madison, Madison, Wisconsin 53706, USA}
\affiliation{Kavli Institute for Theoretical Physics, University of California, Santa Barbara, CA 93106, USA}
\author{Roman M. Lutchyn}
\affiliation{Station Q, Microsoft Research, Santa Barbara, California 93106-6105, USA}

\date{\today}

\begin{abstract}
We study properties of a periodically driven system coupled to a thermal bath. As a nontrivial example, we consider periodically driven metallic system coupled to
a superconducting bath. The effect of the superconductor on the driven system is two-fold: it  (a) modifies density of states in the metal via the proximity effect and
(b) acts as a thermal bath for light-excited quasi-particles. Using Keldysh formalism, we calculate, nonpertubatively in the system-bath coupling, the steady-state properties of
the system and obtain non-equilibrium distribution function. The latter allows one to calculate observable quantities which can be spectroscopically measured in tunneling experiments.
\end{abstract}

\maketitle

\section{Introduction}

The prospects of engineering exotic quantum states of matter using time-periodic driving generated recently much excitement
in condensed matter and cold atom communities~\cite{inoue10,lindner11,kitagawaGF11,Dahlhaus11,jiang11,reynoso12,Liu13,Kundu13,
IadecolaPRL13,Fregoso13,Fregoso14,Iadecola14,Usaj14,FoaTorres14,Sedrakyan15, KitagawaNC12, RechtsmanNature, Struck13, StruckNP13}. The application of these ideas is especially useful
in the context of topological states of matter which are rare in nature.
Thus, the ability of generating various effective time-dependent Hamiltonians is very intriguing because one could engineer topological states using, for example, light-matter interactions.

Periodic driving of isolated non-interacting quantum systems can be understood within the framework of the Floquet formalism which is based on the discrete time-translational invariance of
the Hamiltonian. i.e. $H(t)=H(t+\tau)$ with $\tau$ being the period of driving. Therefore, it is convenient to define the Floquet operator $H_F=H(t)-i\partial_t$~\cite{shirley65,sambe73},
the quasi-energy spectrum of which provides key information about the driven isolated system, see, e.g., Refs.~\cite{Manakov86,Grifoni98}. By suitably engineering time-dependent interaction,
the quasi-energy spectrum may be quite different from the energy spectrum of an equilibrium system~\cite{FloquetReview}. For example, there are current experimental efforts to realize
Floquet topological insulators using the conventional (non-topological) band insulators~\cite{RechtsmanNature13,WangScience13}.

The Floquet formalism describes well the short-time dynamics of the driven system. However, in order to understand the steady-state properties one needs to take into
account system-bath interactions resulting in the relaxation and the redistribution of the Floquet states. Thus, ultimately one needs non-equilibrium  distribution
function in order to understand physical properties of the driven system in the steady-state. The goal of this paper is to address this issue.

Understanding the statistical properties of the periodically driven systems with dissipation is a long-standing problem. Previous studies based on Markovian master equation
formalism adapted for Floquet states indicate that the occupation distribution of Floquet states has rather nontrivial behavior~\cite{Kohler97,Hone97,Breuer00,Kohn01,Hone09,Ketzmerick10}:
the distribution function exhibits Boltzmann-like behavior in some energy range of the spectrum and becomes almost flat in some other intervals.
There have been many attempts to understand steady-state properties in certain special cases. The driven systems with certain symmetries of time-dependent Lagrangian
coupled to a bosonic thermal bath was considered in Refs.~\cite{Iadecola13, LiuFSM15}. The problem of a driven system coupled to a fermion thermal bath has been discussed in Ref.~\cite{Seetharam15}
where it has been shown that Floquet topological systems do not have generically exponential protection against thermal excitations and thermal bath engineering is necessary.
The equilibration and thermalization aspects of Floquet systems have been discussed in Ref.~\cite{IadecolaGCE15}.

Most of the efforts mentioned above are based on Markovian master equation formalism, which relies on the presence of a large time scale separation
(i.e. bath correlation time is much smaller than system relaxation time, and the time scale associated with the driven system dynamics is smaller than
the system relaxation time~\cite{OpenQuantumBook}). Those approximations are valid for weak system-bath coupling (i.e. weak thermalization)~\cite{Hone09}.
If one is interested in the long-time dynamics of a Floquet system strongly coupled to the thermal bath, Born-Markov approximation breaks down and a different
approach is needed. This fact motivates us to look at the problem from a different perspective, and to develop Keldysh formalism which is well-suited for the
problem at hand. This method has been used widely in the context of ac-driven semiconductors (e.g. zero resistance state  phenomenon~\cite{ManiNature02,ZudovPRL03,YangPRL03,AndreevPRL03,Vavilov04,Finkler09})~,
non-equilibrium  superconductivity~\cite{Wyatt66,Wyatt66,Eliashberg70,Robertson09,Mankowsky14,Galitskii69,Elesin71,Galitskii73,Elesin73,AslamazovJETP82,GoldsteinPRB15},
non-equilibrium dynamical mean-field theory~\cite{SchmidtNDMFT02,Freericks06,NDMFT-RMP},
and dissipative system with Lindblad forms~\cite{TorrePRA13,SiebererPRL13,SiebererPRB14,AltmanPRX15,Sieberer15,MaghrebiPRB16}.

In order to demonstrate how this method works we consider a ``toy problem" -- periodically driven metallic system coupled to a fermionic bath, and
calculate steady-state distribution function for arbitrary strength of the system-bath coupling. Next, we consider a more complicated example -- Floquet system
coupled to a superconducting bath which might be relevant for the realization of Floquet counterpart of topological superconductivity~\cite{Fu&Kane08,Sau10,AliceaPRB10,LutchynPRL10,1DwiresOreg}.
For simplicity we do not study this problem in this work since it is straightforward to generalize our method to topological insulators and superconductors.
The fermionic bath, unlike the bosonic one, allows for the quasiparticle exchange (in addition to the energy exchange) which adds certain interesting aspects to the problem that are absent in the bosonic bath case.

From the technical perspective, we use Keldysh Green's function approach for periodically driven systems. We first integrate out the fermionic bath and
incorporate its effects through the self-energy. This allows one to obtain a non-equilibrium  Green's function for the driven system as well as to calculate observable
quantities such as linear differential conductance. The equations for the non-equilibrium  Green's functions now depend on Floquet band indices and therefore become
infinite dimensional. However, analytical solution can be obtained in the limit of small driving amplitude $K$, i.e. $\kappa\equiv K/\Omega\ll 1$ with $\Omega$ being the
driving frequency. For practical reasons, this assumption is not very restrictive since the typical driving frequency $\Omega\sim 1$eV~\cite{RechtsmanNature13,WangScience13}.

The paper is organized as follows. In Sec. \ref{sec:Model}, we introduce our model consisting of a driven metallic system coupled to a fermionic bath.
In Sec. \ref{sec:Keldyshformalism}, we develop a Keldysh formalism for periodically driven system with a bath, and obtain system of equations for the Green's function equations.
Approximate solution for the Green's function in the limit of small driving amplitude is developed in Sec. \ref{sec:Keldyshformalism}. In Secs. \ref{sec:tunnelingS} and \ref{sec:Results},
we discuss non-equilibrium distribution function and physical observables such differential conductance. Finally, we summarize our results in Sec. \ref{sec:conclusions}.


\section{Model for a periodically driven system with the fermionic bath}\label{sec:Model}

We consider a driven metallic system with time-periodic chemical potential, which is coupled to a fermionic reservoir (either normal metal or superconductor). The Hamiltonian of the whole system can be written as
\begin{align}
&H(t)=H_D+H_{{\rm bath}}+H_T,\label{eq:ModelH}\\
&H_D=\sum_{k \sigma}\epsilon_k c_{k\sigma}^{\dagger}c_{k\sigma}+(\mu_{0}+\mu(t))\sum_{k\sigma}c_{k\sigma}^{\dagger}c_{k\sigma},\nonumber\\
&H_T= W\sum_{kq\sigma}(c_{k\sigma}^{\dagger}a_{q\sigma}+h.c.).\nonumber
\end{align}
The Hamiltonian $H_D$ describes the non-interacting driven system with the chemical potential $\mu_{0}+\mu(t)$ where $\mu(t)=\mu(t+\tau)$ and $\tau=2\pi/\Omega$ (we assume $\int_{0}^{\tau}dt\mu(t)=0$). $H_T$ is the tunneling Hamiltonian between the driven system and the
bath. The bath Hamiltonian is given by the mean field BCS Hamiltonian
\begin{equation}\label{eq:H-bath}
 H_{\rm bath}=\sum_{q\sigma} \varepsilon_q a_{q\sigma}^{\dagger}a_{q\sigma}
    +\sum_q (\Delta a_{q\uparrow}^{\dagger}a_{-q\downarrow}^{\dagger}+h.c.)
\end{equation}
where $\Delta$ is $s$-wave pairing potential. The results for the normal metal bath can be obtained by simply setting $\Delta=0$.

It is convenient to study the problem in a rotating frame by applying a time-dependent
unitary transformation $U_{F}(t)=e^{-if(t)\sum_{k\sigma}c_{k\sigma}^{\dagger}c_{k\sigma}}$
with $df(t)/dt=-\mu(t)$, and the Hamiltonian in the rotating frame
becomes
\begin{align}
& H_{F}(t) \!=\! U_{F}^{\dagger}(t)\Big(H(t)-i\partial_{t}\Big)U_{F}(t) \label{eq:H_rotatingF}\\
& \!=\! \sum_{k \sigma}(\epsilon_k\!-\!\mu) c_{k\sigma}^{\dagger}c_{k\sigma}\!+\!H_{{\rm bath}}\!+\!W\sum_{kq}(e^{if(t)} c_{k\sigma}^{\dagger}a_{q\sigma}\!+\!h.c.).\nonumber
\end{align}
In this rotating frame, the time-periodic part of the chemical potential vanishes, and the coupling between the driven system and the
fermionic bath becomes time-dependent. Physically, this term represents photon-induced transitions of quasi-particles between the driven system and
the bath. We will study the interplay between these transitions and the dissipation due to the fermionic bath.

It is convenient to consider the problem using a Keldysh path-integral formalism~\cite{KamenevRev}, where the action on the Keldysh contour can be written as
\begin{eqnarray}
S & = & \sum_{k}\int_{c}dt\int_{c}dt'\overrightarrow{\Psi}{}_{0k}^{\dagger}(t)\breve{Q}_{0}^{-1}(t-t')\overrightarrow{\Psi}{}_{0k}(t')\nonumber\\
 && +\sum_{q}\int_{c}dt\int_{c}dt'\overrightarrow{\Psi}{}_{\rm{bath},q}^{\dagger}(t)\breve{Q}_{\rm{bath},q}^{-1}(t-t')\overrightarrow{\Psi}{}_{\rm{bath},q}(t')\nonumber\\
 && +\sum_{k,q}\int_{c}dt\Big(\overrightarrow{\Psi}{}_{0k}^{\dagger}(t)\breve{M}(t)\overrightarrow{\Psi}{}_{\rm{bath},q}(t)+c.c.\Big).
 \label{eq:action}
\end{eqnarray}
Here we introduced Nambu spinors $\overrightarrow{\Psi}_{0k}^{\dagger}=(c_{k\uparrow}^{\dagger},c_{-k\downarrow})$ and $\overrightarrow{\Psi}_{\rm{bath},q}^{\dagger}=(a_{q\uparrow}^{\dagger},a_{-q\downarrow})$.
The Green's function for the driven system in Nambu space reads
\begin{equation}
\breve{Q}_{0k}^{R}(\omega)=\left(\begin{array}{cc}
\frac{1}{\omega-\epsilon_{k}+i\eta} & 0\\
0 & \frac{1}{\omega+\epsilon_k+i\eta}
\end{array}\right)=\left(\breve{Q}_{0k}^{A}(\omega)\right)^{\dagger}.
\end{equation}
The matrix $\breve{M}$ describes the coupling between $\overrightarrow{\Psi}{}_{0k}^{\dagger}$
and $\overrightarrow{\Psi}_{\rm{bath},q}^{\dagger}$,
and is defined as
\begin{equation}
\breve{M}(t)=\left(\begin{array}{cc}
We^{if(t)} & 0\\
0 & -W^{*}e^{-if(t)}
\end{array}\right).
\end{equation}
We can now integrate out the bath degrees of freedom to find the following effective action:
\begin{equation}
S_{\mathrm{eff}}=\sum_{k}\int_{c}dt\int_{c}dt'\overrightarrow{\Psi}{}_{0k}^{\dagger}(t)\breve{Q}_{k}^{-1}(t,t')\overrightarrow{\Psi}{}_{0k}(t').
\end{equation}
After rewriting the action in terms of the forward
and backward components of the Keldysh contour and performing Larkin-Ovchinnikov rotation (see Ref. [\onlinecite{KamenevRev}]), the Green's function acquires the following matrix form
\begin{equation}
G=\left(\begin{array}{cc} G^R &G^K\\ 0 & G^A \end{array} \right).
\end{equation}
The Dyson's equation for the dressed Green's function becomes
\begin{eqnarray}\label{eq:Qt}
&&\breve{Q}_{k}(t,t')  = \breve{Q}_{0k}(t-t')\nonumber\\
&& +\int_{-\infty}^{\infty}dt_{1}\int_{-\infty}^{\infty}dt_{2}\breve{Q}_{k0}(t-t_{1})\Sigma_{k}(t_{1},t_{2})\breve{Q}_{k}(t_{2},t'),\label{eq:Qk_self}\\
&&\Sigma_{k}(t_{1},t_{2}) = \sum_{q}\breve{M}(t_{1})\breve{Q}_{\rm{bath},q}(t_{1}-t_{2})\breve{M}(t_{2})^{*}
\end{eqnarray}
where $\Sigma_{k}(t_{1},t_{2})$ is the bath self-energy.


\section{Keldysh formalism in the Floquet representation}\label{sec:Keldyshformalism}

As follows from the discussion in the previous section, the Green's function $\breve{Q}_{k}(t,t')$, defined in Eq.~\eqref{eq:Qt}, has two independent time arguments
due to the periodic driving encoded in $\breve{M}(t)$, which breaks continuous time-translational symmetry and only has the discrete symmetry. So the Green's function has the following property $Q(t,t')=Q(t+\tau,t'+\tau)$.
It is convenient to introduce new variables $s=t$ and $u=t-t'$ and define new function $Q(t,t') \rightarrow Q(s,u)$, which satisfies the relation $Q(s,u)=Q(s+\tau,u)$ for all $u$.
One can now perform the following Fourier transformations for $u$
\begin{align}\label{eq:FT1}
 Q(s,\omega)=\int_{-\infty}^{\infty}du e^{-i\omega u}Q(s,u),
\end{align}
and Fourier expansion for $s$:
\begin{align}\label{eq:FT2}
Q(n,\omega)=\frac{1}{\tau}\int_0^{\tau} e^{-i n\Omega s} Q(s,\omega).
\end{align}

Using these identities, we will now derive Dyson's equation in the frequency domain for the problem at hand.
After applying $\int_{-\infty}^{\infty}dt'e^{-i\omega(t-t')}$ and $\frac{1}{\tau}\int_{0}^{\tau}ds\, e^{-in\Omega s}$ to both side of Eq. (\ref{eq:Qk_self}), one finds
\begin{align}
\breve{Q}_{k}(n,\omega)&=\delta_{n0}\breve{Q}_{0k}(\omega)+\sum_{n_{1}}\breve{Q}_{0k}(\omega+n\Omega)\nonumber\\
  &\Sigma_{k}(n_{1},\omega+(n-n_{1})\Omega)\breve{Q}_{k}(n-n_{1},\omega).
  \label{eq:FK_Qself}
\end{align}

We now calculate the self-energy due to the coupling to fermionic bath. The self-energy is \begin{equation}
\Sigma_{k}(t_{1},t_{2})=\sum_{q}\breve{M}(t_{1})\breve{Q}_{\rm{bath},q}(t_{1}-t_{2})\breve{M}(t_{2})^{*},
\end{equation}
which depends on the tunneling matrix
\begin{equation}
\breve{M}(t)  =  \left(\begin{array}{cc}
We^{if(t)} & 0\\
0 & -W^{*}e^{-if(t)}
\end{array}\right)=\sum_{n}e^{in\Omega t}\breve{M}_{n},
\end{equation}
note $df(t)/dt=-\mu(t)$.
In the frequency domain, the self-energy becomes
\begin{equation}
\Sigma_{k}(n,\omega)=\sum_{q}\sum_{n_{2}}\breve{M}_{n+n_{2}}\breve{Q}_{\rm{bath},q}(\omega-n_{2}\Omega)\breve{M}_{n_{2}}^{*}.
\label{eq:FK_Self}
\end{equation}
Combining Eq. (\ref{eq:FK_Qself}) and (\ref{eq:FK_Self}), we obtain a set of coupled equations
\begin{eqnarray}
&&\breve{Q}_{k}(n,\omega)=\delta_{n0}\breve{Q}_{0k}(\omega)+\sum_{n_{1}n_{2}}\breve{Q}_{0k}(\omega+n\Omega)\breve{M}_{n_{1}}\nonumber\\
&&\times\breve{q}_{\rm{bath}}(\omega+(n-n_{1})\Omega)\breve{M}_{n_{2}}^{*}\breve{Q}_{k}(n-n_{1}+n_{2},\omega).
\label{eq:FK_Q_self22}
\end{eqnarray}
where $\breve{q}_{\rm{bath}}(\omega)$ is defined as
\begin{equation}
\breve{q}_{\rm{bath}}(\omega)=\sum_{q}\breve{Q}_{\rm{bath},q}(\omega).
\label{eq:GFbath1}
\end{equation}

Having obtained the self-energy due to the bath, we can calculate the non-equilibrium  Green's function and the distribution function $F(t,t')$ for the driven system.
In general, the relation can be written as
\begin{eqnarray}
\breve{Q}_{k}^{K}(t,t')&=&\int dt_{1}\breve{Q}_{k}^{R}(t,t_{1})F(t_{1},t')\nonumber\\
           &&-\int dt_{1}F(t,t_{1})\breve{Q}_{k}^{A}(t_{1},t')
\end{eqnarray}
After performing the Fourier transform using Eqs. \eqref{eq:FT1} and \eqref{eq:FT2}, one finds
\begin{align}
&\breve{Q}_{k}^{K}(n,\omega)=\sum_{n_{1}}\breve{Q}_{k}^{R}(n_{1},\omega+(n-n_{1})\Omega)F(n-n_{1},\omega)\nonumber\\
      & -\sum_{n_{1}}F(n_{1},\omega+(n-n_{1})\Omega)\breve{Q}_{k}^{A}(n-n_{1},\omega).\label{eq:FK_QKF}
\end{align}

The Green's function of the driven system [i.e. Eq. (\ref{eq:FK_Qself}) and (\ref{eq:FK_QKF})] can be written in the matrix form ($\infty-$ dimension in the Floquet space):
\begin{eqnarray}
\underline{Q_{k}} & = & \underline{Q_{0k}}+\underline{Q_{0k}}\cdot\underline{\Sigma_{k}}\cdot\underline{Q_{k}},\label{eq:MatrixEq_Self}\\
\underline{Q_{k}^{K}} & = & \underline{Q_{k}^{R}}\cdot\underline{F}-\underline{F}\cdot\underline{Q_{k}^{A}}.\label{eq:MatrixEq_KF}
\end{eqnarray}
where the matrices $\breve{Q}_{k}^{\alpha}$, $\Sigma_{k}^{\alpha}$, and
$F$ have the following structure
\begin{equation}
\underline{A}=\left(\begin{array}{ccccc}
\ddots & \cdots & \cdots & \cdots & \ddots\\
\cdots & A(0,\omega+\Omega) & A(1,\omega) & A(2,\omega-\Omega) & \cdots\\
\cdots & A(-1,\omega+\Omega) & A(0,\omega) & A(1,\omega-\Omega) & \cdots\\
\cdots & A(-2,\omega+\Omega) & A(-1,\omega) & A(0,\omega-\Omega) & \cdots\\
\ddots & \cdots & \cdots & \cdots & \ddots
\end{array}\right),\label{eq:AMatrix}
\end{equation}
and the matrix form for $\breve{Q}_{0k}$ is given by
\begin{equation}
\underline{Q_{0k}}=\left(\begin{array}{ccccc}
\ddots & \cdots & \cdots & \cdots & \ddots\\
\cdots & \breve{Q}_{0k}(\omega+\Omega) & 0 & 0 & \cdots\\
\cdots & 0 & \breve{Q}_{0k}(\omega) & 0 & \cdots\\
\cdots & 0 & 0 & \breve{Q}_{0k}(\omega-\Omega) & \cdots\\
\ddots & \cdots & \cdots & \cdots & \ddots
\end{array}\right).\label{eq:Q0Matrix}
\end{equation}
Equations ~\eqref{eq:MatrixEq_Self} and \eqref{eq:MatrixEq_KF} are the main results of this section. Using these results we can calculate density of states as well as occupation distributions
for different Floquet bands by simply solving the matrix equations. This is still a highly nontrivial problem since these matrices are infinite dimensional.
However, controllable analytical solution can be obtained perturbatively in the limit of small driving amplitude $K$ as compared to the driving frequency $\Omega$,
namely in the parameter $\kappa = K/\Omega \ll 1$. In this case, one may truncate the matrix. In the rest of the section, we present our analytical results up to the second order in $\kappa$.
Higher order corrections can be obtained numerically.

For concreteness we consider specific time-dependent perturbation in the form $\mu(t)=-K\cos(\Omega t)$. The corresponding function $f(t)=(K/\Omega)\sin(\Omega t)$, and, thus,  the matrix $M$ can be written as
\begin{equation}
\breve{M}_{n} =  \left(\begin{array}{cc}
W\, J_{n}(\frac{K}{\Omega}) & 0\\
0 & -W^{*}(-1)^{n}\, J_{n}(\frac{K}{\Omega})
\end{array}\right).
\end{equation}
where $J_{n}(x)$ is the Bessel function of the first kind. In the limit of weak driving amplitude, i.e. $\kappa=K/\Omega\ll1$, one can expand above equations up to the lowest order in $\kappa$.
One can show that up to $O(\kappa^2)$, the nonzero matrix elements are for $n=0,\pm1$ and the corresponding matrices are
\begin{eqnarray}
&&\breve{M}_{0}=\left(\begin{array}{cc}
W\, & 0\\
0 & -W^{*}
\end{array}\right),\label{eq:M0}\\
&&\breve{M}_{1}=\left(\begin{array}{cc}
\frac{W}{2}\kappa & 0\\
0 & \frac{W^{*}}{2}\kappa
\end{array}\right),\\
&&\breve{M}_{-1}=\left(\begin{array}{cc}
-\frac{W}{2}\kappa & 0\\
0 & -\frac{W^{*}}{2}\kappa
\end{array}\right)
\end{eqnarray}

This simplification allows one to calculate the non-equilibrium  Green's function explicitly. Using Eq. \eqref{eq:FK_Q_self22}, one finds that the Green's function up to the leading
non-vanishing order in $\kappa$ is given by (see Appendix \ref{app:Greenpert} for details)
\begin{widetext}
\begin{eqnarray}
\breve{Q}_{k}(0,\omega) & = & \frac{1}{\breve{Q}_{0k}(\omega)^{-1}-\breve{M}_{0}\breve{q}_{\rm{bath}}(\omega)\breve{M}_{0}^{*}},\label{eq:Qkband0_1Opert} \label{eq:Q0pert}\\
\breve{Q}_{k}(1,\omega) & = & \frac{1}{\breve{Q}_{0k}(\omega+\Omega)^{-1}-\breve{M}_{0}\breve{q}_{\rm{bath}}(\omega+\Omega)\breve{M}_{0}^{*}}\left(\breve{M}_{0}\breve{q}_{\rm{bath}}(\omega+\Omega)\breve{M}_{-1}^{*}+\breve{M}_{1}\breve{q}_{\rm{bath}}(\omega)\breve{M}_{0}^{*}\right)\nonumber\\
 &  & \qquad\qquad\qquad\qquad\qquad\qquad\times\frac{1}{\breve{Q}_{0k}(\omega)^{-1}-\breve{M}_{0}\breve{q}_{\rm{bath}}(\omega)\breve{M}_{0}^{*}},\label{eq:Qkband1_1Opert} \label{eq:Q1pert}\\
\breve{Q}_{k}(-1,\omega) & = & \frac{1}{\breve{Q}_{0k}(\omega-\Omega)^{-1}-\breve{M}_{0}\breve{q}_{\rm{bath}}(\omega-\Omega)\breve{M}_{0}^{*}}\left(\breve{M}_{0}\breve{q}_{\rm{bath}}(\omega-\Omega)\breve{M}_{1}^{*}+\breve{M}_{-1}\breve{q}_{\rm{bath}}(\omega)\breve{M}_{0}^{*}\right)\nonumber\\
 &  & \qquad\qquad\qquad\qquad\qquad\qquad\times\frac{1}{\breve{Q}_{0k}(\omega)^{-1}-\breve{M}_{0}\breve{q}_{\rm{bath}}(\omega)\breve{M}_{0}^{*}}.\label{eq:Qn1pert}
\end{eqnarray}
To leading order in $\kappa$, the Green's function for the zero Floquet band $\breve{Q}_{k}(0,\omega)$ is the same as the one for a system in the absence of the driving. The corrections to $\breve{Q}_{k}(0,\omega)$ appear in the second order in $\kappa$ and the modified Green's function
$\breve{Q}_{k}(0,\omega)$ becomes
\begin{eqnarray}
\breve{Q}_{k}(0,\omega) & = & \frac{1}{\breve{Q}_{0k}(\omega)^{-1}-\breve{\tilde{M}}_{0}\breve{q}_{\rm{bath}}(\omega)\breve{\tilde{M}}_{0}^{*}-\breve{M}_{1}\breve{q}_{\rm{bath}}(\omega-\Omega)\breve{M}_{1}^{*}-\breve{M}_{-1}\breve{q}_{\rm{bath}}(\omega+\Omega)\breve{M}_{-1}^{*}}\nonumber\\
 &  & +\frac{1}{\breve{Q}_{0k}(\omega)^{-1}-\breve{M}_{0}\breve{q}_{\rm{bath}}(\omega)\breve{M}_{0}^{*}}
      \left(\breve{M}_{0}\breve{q}_{\rm{bath}}(\omega)\breve{M}_{1}^{*}+\breve{M}_{-1}\breve{q}_{\rm{bath}}(\omega+\Omega)\breve{M}_{0}^{*}\right)
      \frac{1}{\breve{Q}_{0k}(\omega+\Omega)^{-1}-\breve{M}_{0}\breve{q}_{\rm{bath}}(\omega+\Omega)\breve{M}_{0}^{*}}\nonumber\\
&  & \times \left(\breve{M}_{0}\breve{q}_{\rm{bath}}(\omega+\Omega)\breve{M}_{-1}^{*}+\breve{M}_{1}\breve{q}_{\rm{bath}}(\omega)\breve{M}_{0}^{*}\right)
      \frac{1}{\breve{Q}_{0k}(\omega)^{-1}-\breve{M}_{0}\breve{q}_{\rm{bath}}(\omega)\breve{M}_{0}^{*}}\nonumber\\
 &  & +\frac{1}{\breve{Q}_{0k}(\omega)^{-1}-\breve{M}_{0}\breve{q}_{\rm{bath}}(\omega)\breve{M}_{0}^{*}}
  \left(\breve{M}_{0}\breve{q}_{\rm{bath}}(\omega)\breve{M}_{-1}^{*}+\breve{M}_{1}\breve{q}_{\rm{bath}}(\omega-\Omega)\breve{M}_{0}^{*}\right)
  \frac{1}{\breve{Q}_{0k}(\omega-\Omega)^{-1}-\breve{M}_{0}\breve{q}_{\rm{bath}}(\omega-\Omega)\breve{M}_{0}^{*}}\nonumber\\
 &  & \times\left(\breve{M}_{0}\breve{q}_{\rm{bath}}(\omega-\Omega)\breve{M}_{1}^{*}+\breve{M}_{-1}\breve{q}_{\rm{bath}}(\omega)\breve{M}_{0}^{*}\right)
       \frac{1}{\breve{Q}_{0k}(\omega)^{-1}-\breve{M}_{0}\breve{q}_{\rm{bath}}(\omega)\breve{M}_{0}^{*}},
 \label{eq:Qkband0_2Opert}
\end{eqnarray}
\end{widetext}
where $\breve{\tilde{M}}_{0}=\left(\begin{smallmatrix} W(1-\frac{\kappa^2}{4})&0\\0&-W^*(1-\frac{\kappa^2}{4})\end{smallmatrix}\right)$
includes the second order term in $\kappa$, while $\breve{M}_{0}$ only has leading order term as shown in Eq. \eqref{eq:M0}.
We also compute the distribution matrix $F(t,\omega)$ perturbatively, see Appendix \ref{app:Greenpert} for details.


\section{Applications}

\subsection{Tunneling spectroscopy} \label{sec:tunnelingS}

We now discuss tunneling experiment and calculate corresponding current between the driven
system at the probe. Similar experiments were pioneered two decades ago as a unique tool to reconstruct intrinsic low-energy quasiparticle relaxation rates in quantum wires.
Electron disequilibrium in the system was induced by applying DC voltage across the wire \cite{PothierPRL97}.
From the bias voltage scaling of the tunnel current data between the probe and wire, one could infer the inelastic quasiparticle scattering rate in the system.
Thus, such measurements carry direct information about the microscopic relaxation processes. The same technique was later generalized and applied to
study superconducting systems \cite{CrosserPRL01,CrosserPRL06,CrosserPRB08}. Importantly, if the density of states in the tunnel probe is completely characterized,
and energy-resolved spectroscopic experiments can directly measure non-equilibrium  quasiparticle occupations.

For the problem of our interest the full Hamiltonian including the tunnel probe can be written as
\begin{equation}
\mathcal{H}(t)=H(t)+H_{TP}+\widetilde{H}_{T}.
\end{equation}
Here $H(t)$ describes the periodically driven system \eqref{eq:ModelH} with the superconducting bath \eqref{eq:H-bath}, while $H_{TP}$ describes the tunnel probe
\begin{equation}
H_{TP}=\sum_{p,\sigma}(\epsilon_{p}+eV)b_{p \sigma}^{\dagger}b_{p \sigma},
\end{equation}
where DC voltage potential $eV$ applied to the tip was included.  Fermion operator $b_{p \sigma}^{\dagger}$ creates an electron with spin $\sigma$ and momentum $p$ in the tunnel probe and $\widetilde{H}_{T}$ describes the tunnel coupling between the tip and the system
\begin{equation}
\widetilde{H}_{T}=\sum_{kp, \sigma}J_{pk}b_{p \sigma}^{\dagger}c_{k \sigma}+h.c.
\end{equation}
We assume that the coupling $J_{kp}$ is very weak, much smaller than the coupling $W$ between the driven system and the superconducting bath, so we can treat this additional probing coupling as a small perturbation.

In the rotating frame with the transformation $U_{F}(t)$, the fermion operators in the system
have an extra time-dependent phase $c_{k \sigma}\rightarrow c_{k \sigma} e^{-i f(t)}$ as shown in previous section. Therefore, in the rotating frame, the current through the tunnel probe is \cite{Jauho94}
\begin{eqnarray}
&&I_{T}(t) =  -e\left\langle\frac{d(\sum_{p}\sum_{\sigma}b_{p \sigma}^{\dagger}b_{p \sigma})}{dt}\right\rangle \nonumber\\
 && =  \frac{e}{\hbar}\sum_{kp \sigma}\Big[J_{pk} i \langle b_{p \sigma}^{\dagger}(t)c_{k \sigma}(t)\rangle e^{-i f(t)}  \nonumber\\
 &&\quad\quad\quad\quad+ J_{pk}^{*} (-i)\langle c_{k \sigma}^{\dagger}(t)b_{p \sigma}(t)\rangle e^{i f(t)}\Big] \nonumber\\
&& =  \frac{2e}{\hbar}\sum_{kp \sigma} \text{Re} \left[ J_{kp}^{*}G_{Ck \sigma,Bp \sigma}^{<}(t,t)e^{i f(t)}  \right].
\end{eqnarray}
We perform the leading order perturbation expansion with respect to the tunneling Hamiltonian
$\widetilde{H}_{T}$ for the Green's function $G_{Cp \sigma,Bk \sigma}^{<}(t,t)$, and obtain
\begin{eqnarray*}
G_{Cp \sigma,Bk \sigma}^{<}(t,t') & = \int_{-\infty}^{\infty}dt_{1}J_{pk}\;\Big[ G_{k \sigma}^{R}(t,t_{1})\; g_{T,p \sigma}^{<}(t_{1}-t')\nonumber\\
                                 & \quad \quad + G_{k \sigma}^{<}(t,t_{1})\; g_{T,p \sigma}^{A}(t_{1}-t') \Big].
\end{eqnarray*}
Here, the function $g_{T,p \sigma}$, e.g. defined as $g_{T,p \sigma}^{<}(t-t_{1})\equiv i\langle b_{p \sigma}^{\dagger}(t_1)b_{p \sigma}(t)\rangle$, is the free fermion Green's function for the tunnel probe,
and the function $G_{k \sigma}$, defined as $G_{k \sigma}^{<}(t,t_{1})\equiv i\langle c_{k \sigma}^{\dagger}(t_1)c_{k \sigma}(t) \rangle$ and $G_{k \sigma}^{>}(t,t_{1})\equiv - i\langle c_{k \sigma}(t) c_{k \sigma}^{\dagger}(t_1) \rangle$
respectively, is the Green's function for the non-superconducting system including the self-energy contribution from superconducting bath. After the further Fourier transformation for $g_{T}$ and assuming $J_{pk}$ independent of $k$, we obtain
\begin{eqnarray}
&I_{T}(t)=\frac{2e}{\hbar}\text{Re}\Bigg\{ \int dt_{1}\int\frac{d\epsilon}{2\pi}e^{-i\epsilon(t-t_{1})} \sum_{\sigma} e^{if(t)} \nonumber\\
    & \times\Big[\left(\sum_{p}|J_{p}|^{2}g_{T,p \sigma}^{<}(\epsilon)\right)\sum_{k}G_{k \sigma}^{R}(t,t_{1})\nonumber\\
    &   +\left(\sum_{p}|J_{p}|^{2}g_{T,p \sigma}^{A}(\epsilon)\right)\sum_{k}G_{k \sigma}^{<}(t,t_{1})\Big] \Bigg\}.
\end{eqnarray}
where we have
\begin{eqnarray*}
\sum_{p}|J_{p}|^{2}g_{T,p \sigma}^{<}(\epsilon) & = & i\;\Gamma_{T}(\epsilon-eV)\; f(\epsilon-eV),\\
\sum_{p}|J_{p}|^{2}g_{T,p \sigma}^{A}(\epsilon) & = & \frac{i}{2}\;\Gamma_{T}(\epsilon-eV),
\end{eqnarray*}
with $eV$ being the constant voltage-energy offset applied to the tunnel probe;
$\Gamma_{T}(\epsilon)=\pi\sum_{p}|J_{p}|^{2}\delta(\epsilon-\epsilon_{p})$
and $f(\epsilon)=1/(e^{\beta\epsilon}+1)$.
For the leading order perturbative calculation in $H_{T}$, if we consider constant density of states in the tip, then the coupling $\Gamma_{T}(\epsilon-eV)=\Gamma_{T}$, and only $f(\epsilon-eV)$ depends on the
voltage. Therefore, the differential conductance is
\begin{equation}
\frac{dI_{T}(t)}{dV}=-2e\!\!\int\!\! \frac{d\epsilon}{2\pi\hbar} \Gamma_{T}\frac{df(\epsilon-eV)}{dV}\sum_{k\sigma} \text{Im}\left[G_{k\sigma}^{R}(t,\epsilon) e^{i f(t)}\right].
\end{equation}
Next we apply the Fourier expansion $G_{k\sigma}^{>,<}(t,\epsilon)=\sum_{n}e^{in\Omega t}G_{k\sigma}^{>,<}(n,\epsilon)$ and
$e^{if(t)}=\sum_{n}e^{in\Omega t} J_{n}(\kappa)$ and perform an average of the current over a full period
\begin{eqnarray}
&&\left\langle\frac{dI_{T}(t)}{dV}\right\rangle_\tau = \frac{1}{\tau}\int_{0}^{\tau}dt\;\frac{dI_{T}(t)}{dV}\\
 & & = -2e\int\frac{d\epsilon}{2\pi\hbar}\Gamma_{T}\frac{df(\epsilon-eV)}{dV}\sum_{nk\sigma}\text{Im}\left[J_{-n}(\kappa) G_{k\sigma}^{R}(n,\epsilon)\right].\nonumber
\end{eqnarray}
Accounting for the fact that $G(n,\epsilon)\propto \kappa^{n}+O(\kappa^{n+2})$, $J_{0}(\kappa)\approx 1-\kappa^2/4+O(\kappa^4)$, and $J_{\pm 1}(\kappa)\approx \pm \kappa/2$,
we find analytical expression for the tunneling density of states of a driven system with the accuracy up to $\kappa^2$ in the form
\begin{eqnarray}
 \nu(\epsilon)&=&\frac{\Gamma_T}{2\pi}\sum_{k\sigma}\Bigg(-2\text{Im}\left[ G_{k\sigma}^{R}(0,\epsilon)\right] (1-\kappa^2/4) \nonumber\\
      && +\kappa \text{Im} \left(G_{k\sigma}^{R}(1,\epsilon)-G_{k\sigma}^{R}(-1,\epsilon)\right)\Bigg).
      \label{eq:DCtipDOS}
\end{eqnarray}


\subsection{Non-equilibrium distribution function}\label{sec:NED}

Another physically interesting and experimentally measurable quantity is the occupation distribution function in the energy space,
which can be computed by the non-equilibrium  Green's functions form the Keldysh block of the matrix Green's function. In general, the distribution function
characterize population of excited states and is important for understanding the statistical mechanics of the dissipative periodically driven systems.
Fermionic occupation in non-superconducting system can be written as
\begin{eqnarray}
&&n_{\sigma}(t)= \sum_{k}\langle c_{k,\sigma}^{\dagger}(t) c_{k,\sigma}(t)\rangle
  =-i\sum_{k}\breve{G}_{k\sigma}^{<}(t,t) \nonumber\\
 && =  -\frac{i}{2}\sum_{k}\int\frac{d\omega}{2\pi}\sum_{n}e^{in\Omega t}
    \Big(\breve{G}_{k\sigma}^{K}(n,\omega) \nonumber\\
    &&-\breve{G}_{k\sigma}^{R}(n,\omega)+\breve{G}_{k\sigma}^{A}(n,\omega)\Big) \nonumber \\
 && \approx  \int d\omega\Bigg[ n_{\sigma}(0,\omega)- \big( e^{i\Omega t}n_{\sigma}(1,\omega) + c.c.  \big) +\cdots  \Big],\label{eq:FermionicOCC}
 \end{eqnarray}
where we define the occupation distribution that can be obtained in the form
\begin{eqnarray}
n_{\uparrow}(0,\omega) & = & -\sum_{k}\frac{i}{4\pi}\Big(\breve{G}_{k\uparrow}^{K}(0,\omega) \nonumber\\
  && -\breve{G}_{k\uparrow}^{R}(0,\omega)+\breve{G}_{k\uparrow}^{A}(0,\omega)\Big),\\
n_{\uparrow}(1,\omega) & = & \frac{i}{4\pi}\sum_{k}\Big(\breve{G}_{k\uparrow}^{K}(1,\omega-\Omega)  \nonumber\\
   && -\breve{G}_{k\uparrow}^{R}(1,\omega-\Omega) +\breve{G}_{k\uparrow}^{A}(1,\omega-\Omega)\Big).
\end{eqnarray}
We have similar expressions for the spin down.
Here we drop the parts beyond the first Floquet band, and use the relation $\breve{G}_{k}^{K}(-1,\omega+\Omega)=-\breve{G}_{k}^{K}(1,\omega)^{*}$and
$\breve{G}_{k}^{R}(-1,\omega+\Omega)=\breve{G}_{k}^{R}(1,\omega)^{*}$.
If the driving is much faster than the system dynamics,
the fast oscillations $e^{in\Omega t}$ for $n\neq 0$ (i.e. the second term and beyond in Eq.(\ref{eq:FermionicOCC})) will
average to zero.


\begin{figure}[t]
\centering
\includegraphics[width=3.25in,clip]{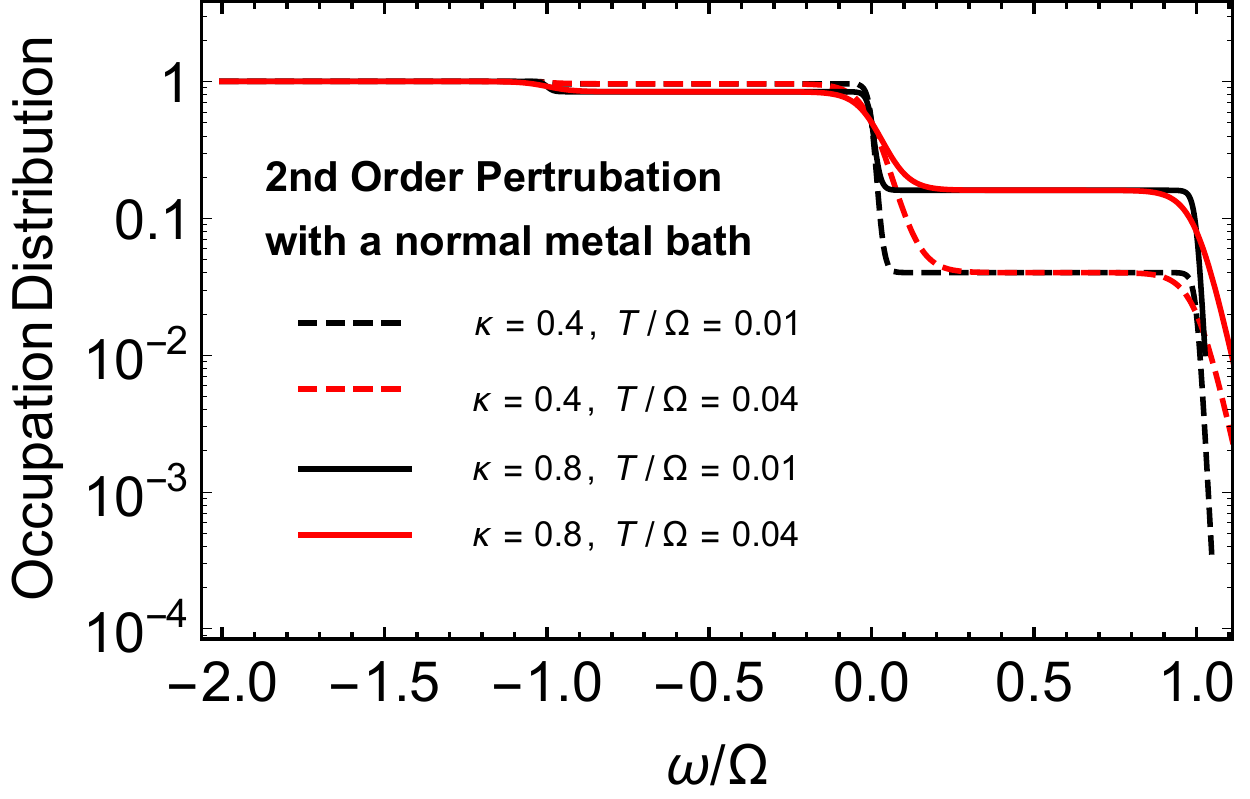}
\caption{Periodically driven system with a normal metal bath. The occupation distribution $n_{\uparrow}(0,\omega)+n_{\downarrow}(0,\omega)$,
computed from the second order perturbation theory for different driving amplitudes $\kappa=K/\Omega=0.4,0.8$, and different temperatures $T/\Omega=0.01,0.04$, is plotted for the zeroth Floquet band.
In the plot we have chosen $W/\Omega=0.08$.}
\label{fig:nonSCbath_2ndOpert}
\end{figure}


\section{Analytical and numerical results} \label{sec:Results}

\subsection{A normal metal bath}
We first consider a normal metal bath where the density of states is constant, i.e. $\rho(\omega)=\rho_F$. After setting $\Delta=0$,
all the Green's functions become diagonal in the Nambu space. Therefore, we restore the conventional fermionic Green's function for normal metals.
In this case, the bath Green's function (in Keldysh space) defined in Eq. (\ref{eq:GFbath1})
can be written as
\begin{eqnarray}
 \breve{q}_{\rm{bath}}&=& \begin{pmatrix}
                         q_{\rm{bath}}^R(\omega) & q_{\rm{bath}}^K(\omega)\\
                         0 & q_{\rm{bath}}^A(\omega)
                        \end{pmatrix}\nonumber\\
                      &=& \begin{pmatrix}
                        -i\pi \rho_F & -2 i\pi \rho_F (1-2 f(\omega) ) \\
                         0  & i\pi \rho_F
                       \end{pmatrix}.
\label{eq:NMbathGreen}
\end{eqnarray}
where $f(\omega)=1/(e^{\omega/T}+1)$ is the Fermi distribution function.
After substituting Eq. (\ref{eq:NMbathGreen}) into Eqs. (\ref{eq:Qkband0_2Opert}), (\ref{eq:Q1pert}) and (\ref{eq:Qn1pert}), the Green's function (computed with the accuracy up to the second order in $\kappa$)
can be simplified for $n=0$ to
\begin{eqnarray}
 &&G_{k\sigma}^R(0,\omega)=G_{k\sigma}^A(0,\omega)^* = \frac{1}{i \Gamma +\omega -\epsilon_k }\label{eq:nonSC_GRAk}   \\
 &&G_{k\sigma}^K(0,\omega) =  \nonumber\\
 &&\frac{i \Gamma  \left(\kappa ^2 (f (\omega -\Omega )+f (\omega +\Omega )-2 f(\omega))+4 f(\omega) -2\right)}{\Gamma ^2
     +(\epsilon_k -\omega )^2},\label{eq:nonSC_GKk}
\end{eqnarray}
and for $n=\pm 1$ to
\begin{eqnarray}
 &&G_{k\sigma}^R(\pm 1,\omega) = 0, G_{k\sigma}^A(\pm 1,\omega) = 0, \\
 &&G_{k\sigma}^K(\pm 1,\omega) = \frac{\pm 2 i \kappa \Gamma \left( f(\omega)-f(\omega\pm\Omega)\right)}
              {\Gamma^2 +(\omega-\epsilon_k )(\omega\pm\Omega-\epsilon_k )\mp i\Gamma \Omega},
\end{eqnarray}
which recovers the Green's functions for the equilibrium case for $\kappa\rightarrow 0$ or $\Omega\rightarrow 0$. Above we have defined $\Gamma=\pi\rho_F W^2$. The retarded and advanced parts
of the Green's function matrix for $n\neq 0$ case are vanishing only if the bath density of states is assumed to be a constant.
As shown in the last section, a normal metal tunnel probe with DC voltage bias can be used to determine the tunneling density of states, which only depends on zero Floquet band ($n=0$) for normal metal bath
\begin{equation}
\nu(\epsilon)\approx -\frac{\Gamma_T}{\pi}\sum_{k\sigma}\text{Im}\left[G_{k\sigma}^{R}(n=0,\epsilon)\right](1-\kappa^2/4).
\end{equation}

\begin{figure}[t]
\centering
\includegraphics[width=3.25in,clip]{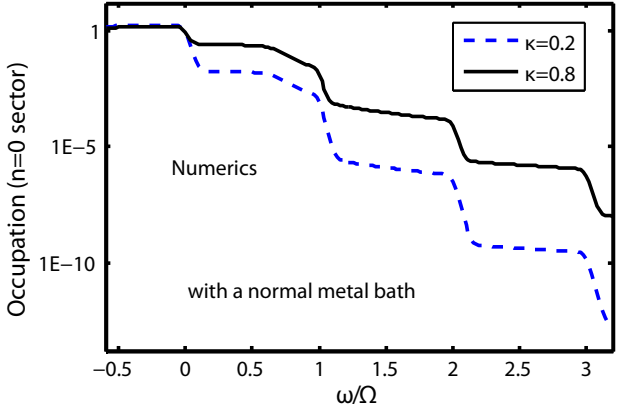}
\caption{Periodically driven system with a normal metal bath.
The occupation distribution $n_{\uparrow}(0,\omega)+n_{\downarrow}(0,\omega)$ is plotted for the zeroth Floquet band using numerically evaluation of Green's function
equations (\ref{eq:MatrixEq_Self}-\ref{eq:Q0Matrix}) for different driving amplitudes $\kappa=0.2$ and $0.8$. We have chosen $V/\Omega=0.08$ and $T/\Omega=0.04$.}
\label{fig:nonSCbath_NUM}
\end{figure}

This non-equilibrium  Green's function directly reveals the information
of the occupation of the system
\begin{eqnarray}
\overline{n_{\sigma}(t)} = \frac{1}{\tau}\int_{\tau}^0 dt n_{\sigma}(t)
 = \int d\omega n_{\sigma}(0,\omega),
 \end{eqnarray}
and their distribution can be obtained
\begin{eqnarray}
&&n_{\sigma}(0,\omega) = -\sum_{k}\frac{i}{4\pi}\Big(\breve{G}_{k\sigma}^{K}(0,\omega)
   -\breve{G}_{k\sigma}^{R}(0,\omega)+\breve{G}_{k\sigma}^{A}(0,\omega)\Big)\nonumber\\
  &&= \frac{1-\tanh\left(\frac{\omega}{2T}\right)}{2} +\frac{ \left(\cosh \left(\frac{\Omega }{T}\right)-1\right)}{8} \tanh \left(\frac{\omega }{2 T}\right)   \nonumber\\
  && \quad\quad\quad\quad\times \text{sech}\left(\frac{\omega -\Omega }{2 T}\right) \text{sech}\left(\frac{\omega +\Omega }{2 T}\right) \kappa ^2 + O(\kappa^3)  \nonumber\\
  &&\xrightarrow{T\rightarrow 0} \begin{cases}
                                  1+ O(\kappa^3) & \quad \text{if  } \omega < -\Omega\\
                                  1-\frac{\kappa^2}{4}+ O(\kappa^3) & \quad \text{if  } -\Omega\leq\omega\leq 0\\
                                  \frac{\kappa^2}{4}+ O(\kappa^3) & \quad \text{if  } 0<\omega<\Omega\\
                                  O(\kappa^3) & \quad \text{if  } \omega\geq\Omega\\
                                 \end{cases}
\end{eqnarray}
We plot this occupation distribution function in Fig. \ref{fig:nonSCbath_2ndOpert}. The interplay between the driving potential
and bath dissipation causes multi-step suppressions and show plateaus between $\omega=0$ and $\omega=\pm\Omega$ in the non-equilibrium  stationary occupation
function. This draws some analogies to observed multi-step-structure of non-equilibrium  steady states as observed in the energy-resolved tunneling experiments with diffusive quantum wires \cite{PothierPRL97}.
In that case, steps occur due to admixture of Fermi distributions in the leads by a voltage bias whereas rounding of steps is governed by inelastic electron-electron collisions in the wire.
We also notice that in the setup considered here the non-equilibrium  population can be controlled by tuning driving amplitude.


\begin{figure}[t]
\centering
\includegraphics[width=3.25in,clip]{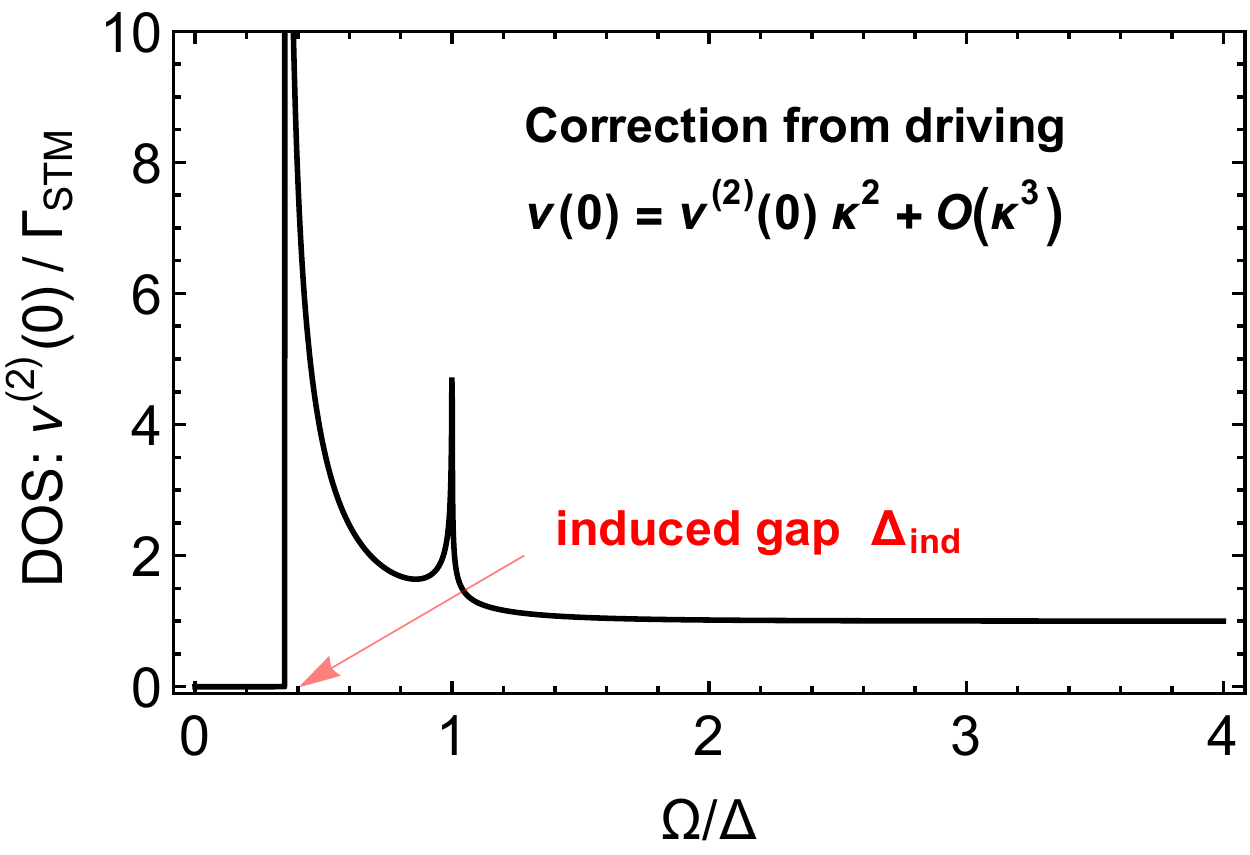}
\caption{The driving-induced correction to the density of states of the first Floquet band at zero frequency $\nu^{(2)}(\omega=0)$ as a
function of the driving frequency. We have chosen following parameters $\rho_F \Delta=1.0$, $W/\Delta=0.4$ (note $\Gamma=\pi W^2 \rho_F$).
}
\label{fig:SCbath_analy}
\end{figure}


Next, we numerically compute the occupation distribution (for zeroth Floquet band) by evaluating the Matrix Eqs. (\ref{eq:MatrixEq_Self}-\ref{eq:Q0Matrix}).
In this calculation, we truncate the matrices and choose a large enough $N$, such that the spectrum becomes unchanged with further increasing $N$.
Here, the driving amplitude $\kappa$ is still smaller than one, we will choose a finite energy band,
i.e. the summation $\sum_k$ is replaced by $\int_{-D}^{D} d \epsilon_k$ with finite $D$, for the non-superconducting system so that the truncation of the matrices is valid and efficient.
We plot numerical results of $\nu(n=0,\omega)$ in Fig. \ref{fig:nonSCbath_NUM} for different driving amplitudes $\kappa=0.2$ and $0.8$.
Numerical results show multi-step suppression and multi-plateau regimes between $\omega=n \Omega$ and $\omega=(n+1) \Omega$
for $n=0,1,2,\cdots$. Similar multi-step staircase electron energy distribution function was predicted to occur in a mesoscopic AC-driven diffusive wire
with the step width controlled by the field energy quantum \cite{ShytovPRB05}. It should be emphasized that generally such distribution is not characterized by an effective temperature.
This rich structure of the distribution can be resolved in tunneling experiments. It also has important consequences for the current shot noise measurements as the current power spectrum of fluctuations is determined
by a spectral integral of the product $n(\omega)[1-n(\omega)]$. In particular, in the shot noise limit one expects multi-step distribution to translate into distinct Fano factor.

\begin{figure}[t]
\centering
\includegraphics[width=3.4in,clip]{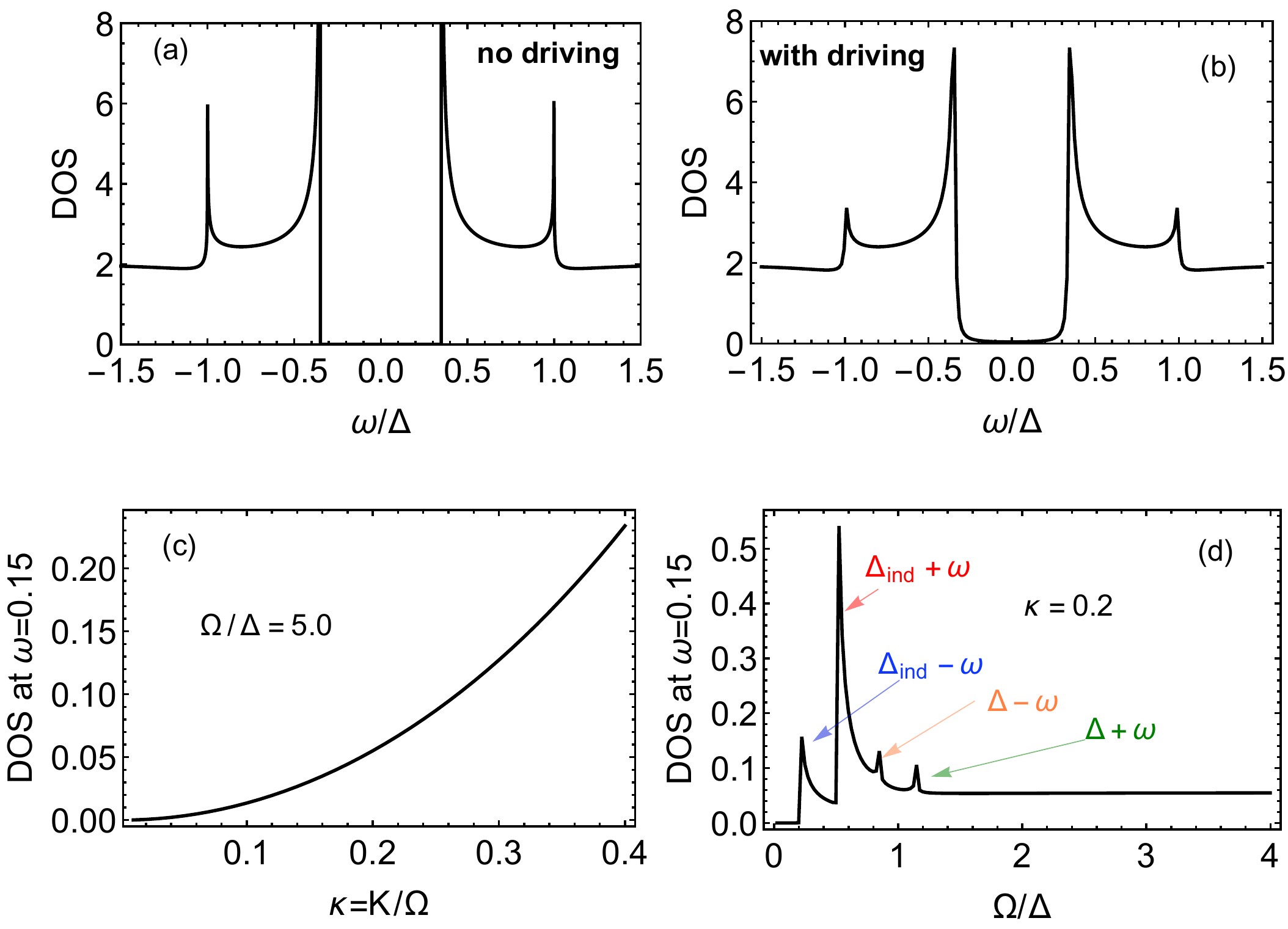}
\caption{The density of states $\nu(\omega)$ for the zeroth Floquet band from the perturbative calculation. (a) the result without driving potential, i.e. the equilibrium result; (b)
the result including the second order in $\kappa$, where we choose coupling between fermionic system and bath $W/\Delta=0.4$, the driving amplitude $K/\Omega=0.2$, and $\Omega/\Delta=5.0.$
The DOS at $\omega/\Delta=0.15$: $\nu(0.15)/\Gamma_{T}$, (c) as
a function of  $\kappa$ for fixed $\Omega/\Delta=5.0$, (d) as a function of $\Omega/\Delta$ for fixed $\kappa=K/\Omega=0.2$.}
\label{fig:DOS_Perturbation}
\end{figure}


\subsection{A superconducting bath}

\subsubsection{Tunneling density of states}

In this section, we consider more interesting case of a superconducting bath, where the bath Green's function
can be expressed in terms of the quasiclassical Green's function for a BCS superconductor~\cite{Eilenberger68,LarkinQC,Serene&Rainer}
\begin{align}
&\breve{q}_{\rm{bath}}(\omega)=\sum_{q}\breve{Q}_{\rm{bath},q}(\omega)\label{eq:GFbath1}\\
&=\rho_{F}\int d\epsilon_{q}\int\frac{d\Omega_{q}}{4\pi}\breve{Q}_{\rm{bath},q}(\omega)= -i\pi\rho_{F}g^{qc}(\omega),\nonumber
\end{align}
The retarded, advanced, and Keldysh components of the quasiclassical Green's functions for an $s$-wave superconductor read
\begin{align}
g^{qc,R/A}(\omega) & =  \frac{1}{\sqrt{(\omega\pm i\eta)^{2}-\Delta^{2}}}\left(\begin{array}{cc}
\omega\pm i\eta & -\Delta\\
-\Delta & \omega\pm i\eta
\end{array}\right),\label{eq:QCGFSC_R}\\
g^{qc,K}(\omega) & =  g^{qc,R}(\omega)h(\omega)-h(\omega)g^{qc,A}(\omega),\label{eq:QCGFSC_K}
\end{align}
where  $\eta \rightarrow 0$; the equilibrium
distribution function is
\begin{equation}
h(\omega)=\left(\begin{array}{cc}
{\rm tanh}\left(\frac{\omega}{2T}\right) & 0\\
0 & {\rm tanh}\left(\frac{\omega}{2T}\right)
\end{array}\right).
\end{equation}
In order to recover the results for a normal metal bath, we can simply set $\Delta=0$.
For a superconductor, the bath density of states is not a constant, thus the retarded and advanced part of the $Q(n\neq 0,\omega)$ are not vanishing.
Consequently, the tunneling density of states with a DC tip has corrections from Green's function with $n\neq 0$, and with the accuracy up to $\kappa^2$ we have to keep the $n=\pm 1$ contribution in Eq. (\ref{eq:DCtipDOS})
\begin{equation}
 \nu(\epsilon)=\nu(0,\epsilon)+\nu(1,\epsilon)+O(\kappa^3)
\end{equation}
where $\nu(1,\epsilon)$ describes $n=\pm 1$ contributions.

Let us first look at the tunneling density of states for the zeroth Floquet band ($n=0$), which is defined as
\begin{eqnarray}
&&\nu(0,\epsilon)/\Gamma_{T} = -\frac{1}{\pi}\sum_{k\sigma}\text{Im}G_{k\sigma}^{R}(n=0,\epsilon) \left(1-\frac{\kappa^2}{4}\right)\\
&&=-\frac{1-\frac{\kappa^2}{4}}{\pi}\text{Im}\left[{\rm Tr}\left[\left(\begin{array}{cc}
1 & 0\\
0 & C^{*}
\end{array}\right)\sum_{k}\breve{Q}_{k}^{R}(0,\omega)\left(\begin{array}{cc}
1 & 0\\
0 & C
\end{array}\right)\right]\right].\nonumber
\end{eqnarray}
where $C$ is the charge conjugation operator. In this case, the analytic result in the $\omega\rightarrow 0$ limit up to second order in $\kappa$ can be simplified. We expand the DOS in small $\kappa$
\begin{equation}
 \nu(0,0)/\Gamma_T=\nu^{(0)}(0,0)+\kappa^2 \nu^{(2)}(0,0)+O(\kappa^3).\label{eq:nuN0}
\end{equation}
For finite $\Delta$, we have $\nu^{(0)}(0,0)=0$ as expected, which corresponds to the DOS for the equilibrium system. We focus on the second part $\nu^{(2)}(0,0)$.
The full analytic expressions are very involved and not enlightening, we refer the reader to Appendix \ref{app:AnalyticExpression}.
Here, we only present some limits that can be simplified for two different regimes: $\Omega<\Delta$ and $\Omega>\Delta$ respectively
\begin{equation}
 \nu^{(2)}(0,0)=
    \begin{cases}
      1   &\;\text{for} \;\Omega\gg\Delta\\
      \mathrm{Re}\left[ \frac{\sqrt{\Gamma }}{2^{3/4} \sqrt[4]{-\Delta } \sqrt[4]{\Omega -\Delta }} \right] &\; \text{for}\; \Omega\rightarrow \Delta+0\\
      \mathrm{Re}\left[ \frac{\sqrt{\Gamma }}{2^{3/4} \sqrt[4]{\Delta } \sqrt[4]{\Delta -\Omega }} \right] &\; \text{for}\; \Omega\rightarrow \Delta-0\\
    \end{cases}
\end{equation}
where we take the limit $\eta\rightarrow 0$.
Secondly, we consider contributions from the higher Floquet band ($n=\pm 1$), which is given by
\begin{eqnarray}
\frac{\nu(1,\omega)}{\Gamma_{T}} &=& \frac{\kappa}{2\pi} \sum_{k\sigma} \text{Im} \left(G_{k\sigma}^{R}(1,\omega)-G_{k\sigma}^{R}(-1,\omega)\right)\nonumber\\
&=&\frac{\kappa}{2\pi} \sum_{k}\text{Im}{\rm Tr}\Bigg[\left(\begin{array}{cc}
1 & 0\\
0 & C^{*}
\end{array}\right)\sum_{k}\Big(\breve{Q}_{k}^{R}(1,\omega)\nonumber\\
&&\quad\quad\quad          -\breve{Q}_{k}^{R}(-1,\omega)\Big)\left(\begin{array}{cc}
1 & 0\\
0 & C
\end{array}\right)\Bigg].
\end{eqnarray}
Again we focus on analytical expressions at $\omega=0$
\begin{equation}
 \nu(1,0)/\Gamma_T = \kappa^2 \nu^{(2)}(1,0)+O(\kappa^3).\label{eq:nuN1}
\end{equation}
The analytical expression for $\nu^{(2)}(1,0)$ is also cumbersome and can be found in appendix \ref{app:AnalyticExpression}. This function can be simplified in certain limits
\begin{equation}
 \nu^{(2)}(1,0)=
    \begin{cases}
      \quad\frac{2\Gamma^2}{\Omega^2}   &\;\text{for} \;\Omega\gg\Delta\\
      \text{Re}\left[ -\frac{\sqrt{-\Delta } \sqrt{\Gamma  \sqrt{-\Delta } \Delta ^2}}{2^{3/4} \Delta ^2 \sqrt[4]{\Omega -\Delta }} \right] &\; \text{for}\; \Omega\rightarrow \Delta+0\\
      \text{Re}\left[ \frac{\sqrt{\Gamma  \Omega ^{5/2}}}{2^{3/4} \Omega ^{3/2} \sqrt[4]{\Delta -\Omega }} \right] &\; \text{for}\; \Omega\rightarrow \Delta-0\\
    \end{cases}
\end{equation}

For completeness, we numerically evaluate the driving-induced correction $\nu^{(2)}(0)=\nu^{(2)}(0,0)+\nu^{(2)}(1,0)$ in Fig. \ref{fig:SCbath_analy}.
To the leading order in $\kappa$, this correction for $\omega=0$ vanishes when driving frequency is smaller than the
superconducting proximity-induced energy gap $\Omega<\Delta_{\rm ind}$. For the arbitrary relation between $\Gamma$ and $\Delta$ this gap has a complicated form.
In the case of weak coupling $\Gamma\ll\Delta$, following asymptotic formula applies $\Delta_{\mathrm{ind}}\approx\Gamma-\Gamma^2/\Delta$.
Because of the resonant transitions between the ground state and the gap edges, the driving-induced correction also exhibits BCS singularities
around $\Omega\sim \Delta_{\rm ind}$ and $\Omega\sim \Delta$. To obtain the analytic expressions, we assume finite $\Delta$ and take the limit $\omega=0$, where the procedure has ambiguity; therefore,
we cannot directly take $\Delta\rightarrow 0$ limit for those expressions to compare with the normal metal bath results.
At $\omega\neq 0$, the boundary of zero DOS regime should be smaller than $\Delta_{\rm ind}$; for the higher order corrections,
the DOS is not exactly vanishing for small driving frequency in the regime $\Omega<\Delta_{\rm{ind}}$.
Physically, those leakage DOS within the induced gap comes from the processes involving higher Floquet bands, while up to the leading order correction in $\kappa$,
only the first lowest bands (both $n=1$ and $n=-1$) are involved.


\begin{figure}[htp]
\centering
\includegraphics[width=3.4in,clip]{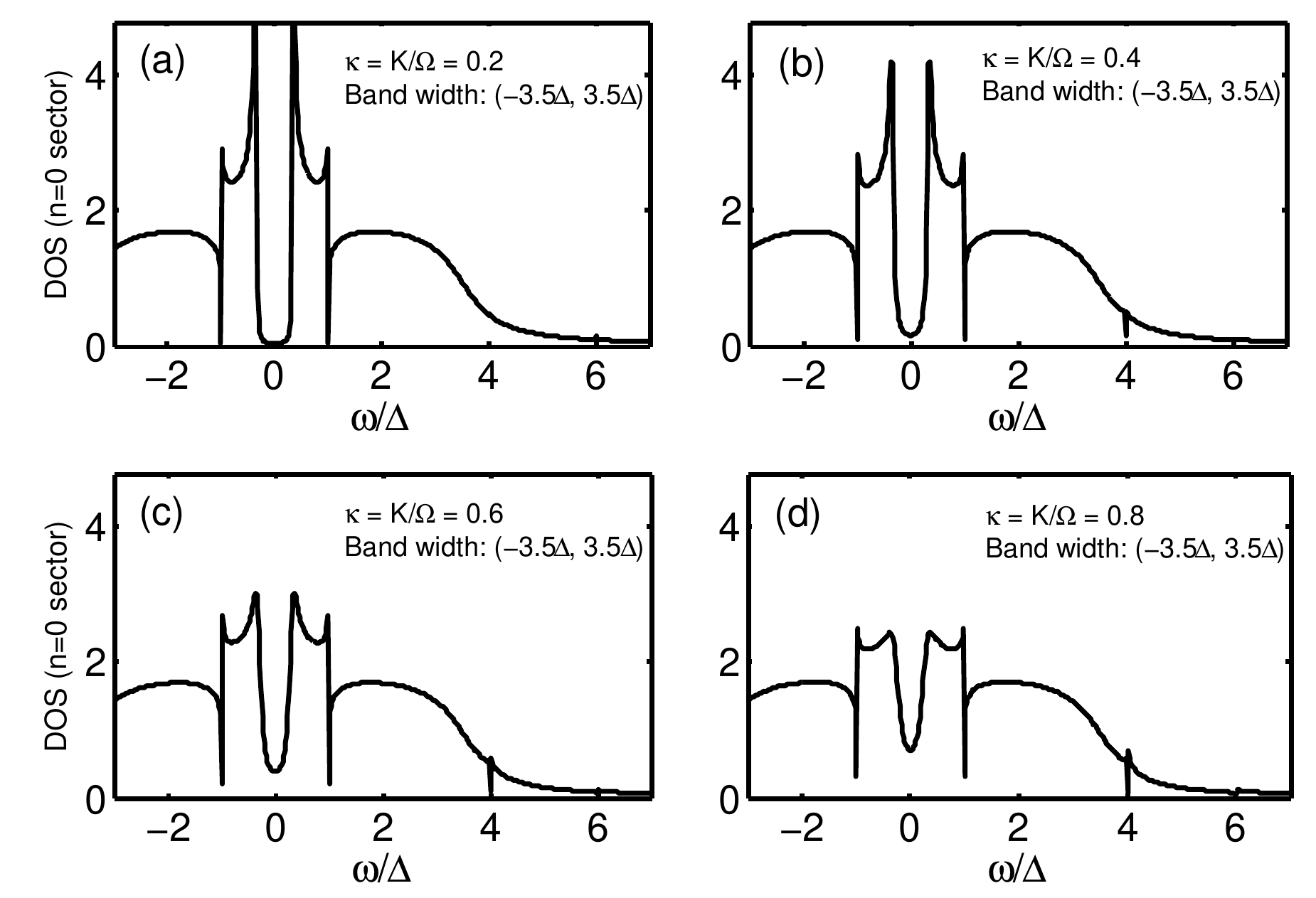}
\caption{The density of states  $\nu(\omega)$ for the zeroth Floquet band from the numerical solution of  Eqs. (\ref{eq:MatrixEq_Self}-\ref{eq:Q0Matrix}). We choose $\Omega/\Delta=5.0$, $W/\Delta=0.4$, and finite band width for the normal metal part $(-D,\, D)$ with $D/\Delta=3.5$, and for different driving amplitude (a) $\kappa=0.2$,(b) $\kappa=0.4$,(c) $\kappa=0.6$, and (d) $\kappa=0.8$.
$N=30$ ($2N+1$ Floquet bands).}
\label{fig:DOSn0_Num}
\end{figure}

\begin{figure}[htp]
\centering
\includegraphics[width=3.25in,clip]{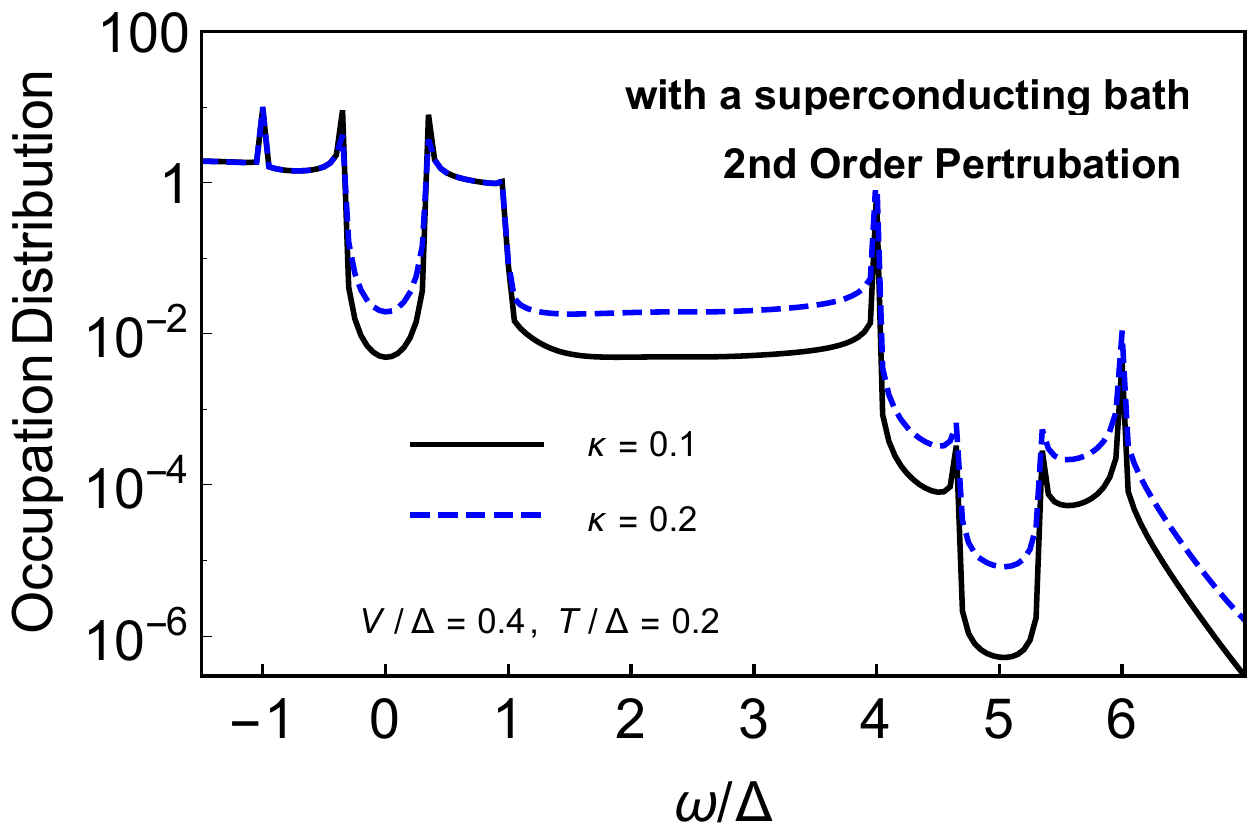}
\caption{ The occupation distribution $n_{\uparrow}(0,\omega)+n_{\downarrow}(0,\omega)$
of the zeroth Floquet band from the second order perturbation calculation in $\kappa$.
We took coupling between fermionic system and bath $W/\Delta=0.4$, the driving amplitudes $\kappa =0.2, 0.1$, $\Omega/\Delta=5.0$ and, $T/\Delta=0.2.$.}
\vspace{-0.2in}
\label{fig:OCC0band_Perturbation}
\end{figure}


The general analytic expressions of the Green's functions and finite energy become very complicated so we will only show the
numerical results for $\omega\neq 0$ and discuss certain asymptotic limits. Results for the zeroth Floquet band computed (i) without driving potential,
i.e. equilibrium case and (ii) with driving potential up to second order in $\kappa$ [Eq. (\ref{eq:Qkband0_2Opert})] are shown in Fig. \ref{fig:DOS_Perturbation}
(a) and (b) respectively. In equilibrium, the DOS within the induced gap is exactly zero. When including the second order correction in $\kappa $, that accounts for the
processes virtually involving other Floquet band ($\pm 1$ band here), one finds leakage of the DOS within the induced gap $(-\Delta_{\mathrm{ind}},\;\Delta_{\mathrm{ind}})$,
which is shown in Fig. \ref{fig:DOS_Perturbation} (b). At smallest energies, $\omega\ll \{\Gamma,\Delta\}\ll\Omega$, the asymptotic expression reads  $\nu^{(2)}(0,\omega)\approx1+(2\Gamma^2+6\Gamma\Delta+3\Delta^2)\omega^2/2\Gamma^2\Delta^2$.
At higher energies there are two power-law singularities in the DOS at induced $\omega\sim\Delta_{\mathrm{ind}}$ and bulk $\omega\sim\Delta$ energy gaps.
We also numerically computed the leakage of DOS within the gap for $\omega\neq 0$, e.g. $\nu(\omega/\Delta=0.15)$,
as a function of $\kappa$ for fixed $\Omega/\Delta=5.0$ [in Fig. \ref{fig:DOS_Perturbation} (c)], and as a function of $\Omega/\Delta$ for fixed $\kappa=K/\Omega=0.2$
[as shown in Fig. \ref{fig:DOS_Perturbation} (d)]. As a function of drive frequency DOS shows sharp peak structure at the $\Omega=\Delta_{\mathrm{ind}}\pm \omega$ and $\Delta\pm \omega$,
and saturates to a constant with further increasing ratio $\Omega/\Delta$.


\begin{figure}[t]
\centering
\vspace{0.1in}
\includegraphics[width=3.25in,clip]{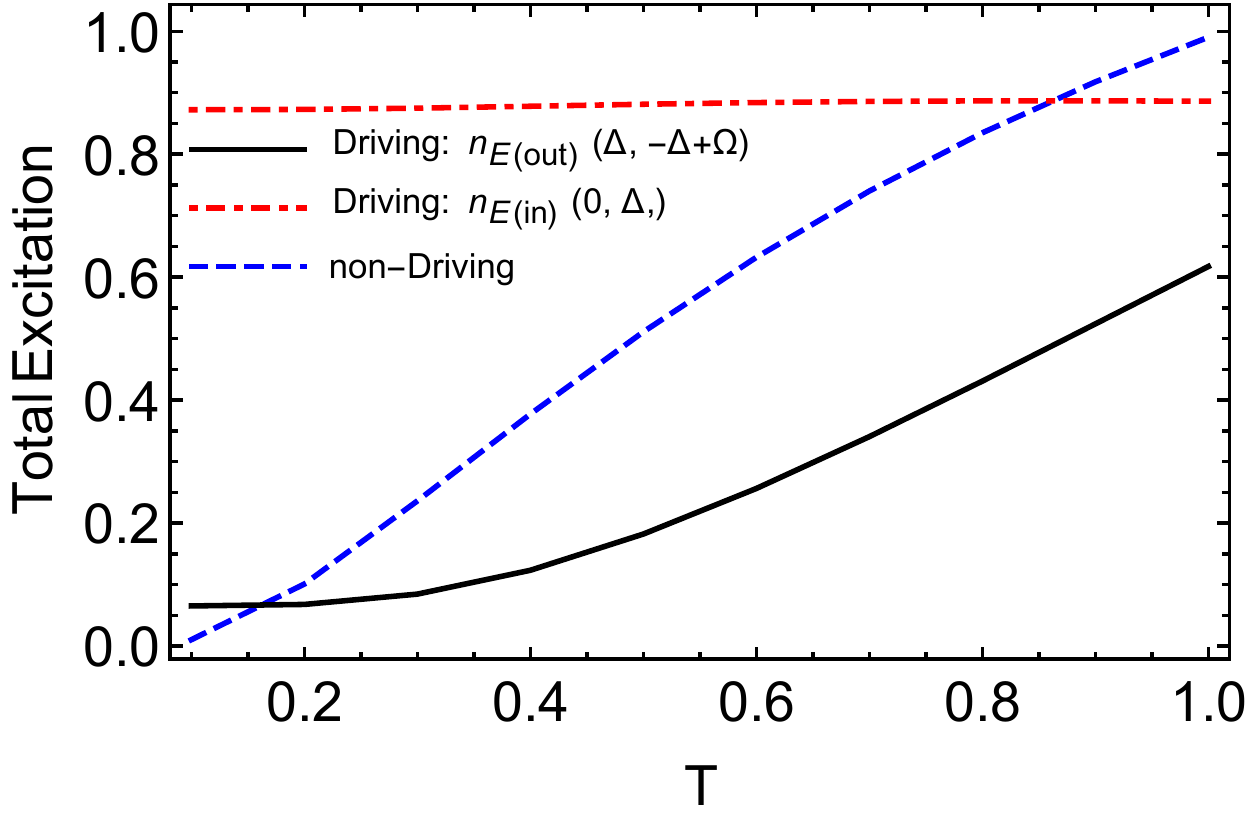}
\caption{ The integrated occupations $n_{0}$ of Eq. (\ref{eq:totExc_Pert}) of the zeroth Floquet band above
$\omega=0$ as a function of temperature obtained from 2nd order perturbation calculation. Black solid line: with the drive; blue dashed line: equilibrium case. For this plot we took coupling between fermionic system
and bath $W/\Delta=0.4$, the driving amplitude $\kappa =0.2$, and $\Omega/\Delta=5.0$.}
\label{fig:totalOCC_Q0}
\end{figure}


Following the same procedures as in the case of normal metal bath and evaluating the full matrices
in Eqs. (\ref{eq:MatrixEq_Self}-\ref{eq:Q0Matrix}), we can obtain the numerical results for DOS $\nu(n=0,\omega)$, see Fig. \ref{fig:DOSn0_Num}.
At small driving amplitude $\kappa=0.2$ full numerical result shown in Fig. \ref{fig:DOSn0_Num} (a) is qualitatively similar to the result of
analytical perturbative calculation. Indeed, one finds leakage of states under the proximity induced gap which becomes progressively more pronounced
with increasing amplitude. Gap is completely lifted at driving exceeding $\kappa\sim0.4$, however DOS remains depleted in the energy window of the
order $\Delta_{\mathrm{ind}}$. With further increasing driving energy states between induced and hard gap tend to fill completely. All these features are clearly visible in Fig. \ref{fig:DOSn0_Num}.

It should be noted that in an earlier study \cite{NarozhnyPRB99} of a normal metal-superconductor systems in the regime,
when electrons in the normal metal were driven out of equilibrium by a DC voltage bias, it was shown that non-equilibrium
fluctuations of the electron density in the N-layer cause the fluctuations of the phase of the order parameter in the S-layer.
As a result, the density of states in the superconductor was shown to deviate from the standard BCS form, the density
of states in the gap becomes finite. This effect was interpreted as a result of the time reversal symmetry breaking due to the non-equilibrium,
and was described in terms of a low energy collective Schmid-Sch\"on-like mode of the junction, which couples normal currents in N-layer and supercurrents in the S-layer.
The broadening of the singularity of the density of states in the S-layer was found to manifest itself similarly to the broadening of the distribution function.
Our findings here, although complimentary, are different in their physical essence. We considered a limit when superconductor has nonperturbative
effect on a spectrum of a normal layer itself inducing an energy gap $\Delta_{\mathrm{ind}}$. At equilibrium DOS of that N-layer is zero below the induced gap,
however periodic drive induces finite occupation of sub-gap states where superconductor serves as a bath mediating dissipative processes and thus stabilizing steady state distributions.

\subsubsection{Occupation distribution and non-equilibrium Green's function}

In this subsection we study the occupation distribution and non-equilibrium Green's function for the system with a superconducting bath. As discussed in Sec. \ref{sec:NED}, fermionic occupation can be written as
\begin{eqnarray}
n_{\uparrow}(t)
  \approx \int d\omega\Bigg[ n_{\uparrow}(0,\omega)- \big( e^{i\Omega t}n_{\uparrow}(1,\omega) + c.c.  \big)  \Big],
\end{eqnarray}
where the corresponding distribution functions are defined as
\begin{eqnarray}
n_{\uparrow}(0,\omega) & = & -\sum_{k}\frac{i}{4\pi}\Big(\breve{Q}_{k,\uparrow\uparrow}^{K}(0,\omega) \nonumber\\
  && -\breve{Q}_{k,\uparrow\uparrow}^{R}(0,\omega)+\breve{Q}_{k,\uparrow\uparrow}^{A}(0,\omega)\Big),\\
n_{\uparrow}(1,\omega) & = & \frac{i}{4\pi}\sum_{k}\Big(\breve{Q}_{k,\uparrow\uparrow}^{K}(1,\omega-\Omega)  \nonumber\\
   && -\breve{Q}_{k,\uparrow\uparrow}^{R}(1,\omega-\Omega) +\breve{Q}_{k,\uparrow\uparrow}^{A}(1,\omega-\Omega)\Big).
\end{eqnarray}
Here $\breve{Q}_{k,\sigma\sigma}$ is the diagonal part of the Nambu Green's function.
The spin down channel has the similar form, and we drop the parts beyond the first Floquet band.
If the driving is much faster than the system dynamics, we can drop the fast oscillation ($e^{in\Omega t}$) parts, i.e. the second term and beyond.

We first consider perturbative expansion. For this case, the analytic results in the zero frequency $\omega=0$ can be simplified, and we find the Keldysh
component in occupation distribution $n_{\sigma}(0,\omega=0)$ is exactly zero, i.e. $\sum_{k}\breve{Q}_{k,\uparrow\uparrow}^{K}(0,\omega=0)=0$.
Therefore, the occupation of the zeroth Floquet band at $\omega=0$ is proportional to the corresponding DOS with a pre-factor $\frac{1}{2}(1-\frac{\kappa^2}{4})$.
We would expect nontrivial results for $\omega>0$, their analytic form becomes very complicated and less transparent. We, therefore, only show numerical results.
The non-oscillatory part $n(0,\omega)$ is shown in Fig. \ref{fig:OCC0band_Perturbation} based on the second order perturbation calculation.
In equilibrium, the small finite temperature effects can only induce a small electron excitation occupation above the induced-gap. However, with periodic driving potential, (i) electrons
can be excited above the induced-gap and has large stationary distribution between the induced-gap and the bath bulk-gap [$\omega/\Delta\sim 0.2$ to $1.0$ as shown in Fig. \ref{fig:OCC0band_Perturbation}],
and the population in this regime is almost independent of the driving amplitude and dissipation. (ii) Within the induced gap, the occupation will be enhanced by increasing driving amplitude due to the leakage of states.
(iii) We also notice that the occupation is significantly reduced for the energy above the bath superconducting gap.
The reason is that the bulk superconductor serves as dissipative bath, but the bath has no DOS below the bulk gap $\Delta$,
therefore, dissipation play little role for $\omega<\Delta$ in cases (i) and (ii), only when electrons are excited above $\Delta$, the bath will interplay significantly with the excited fermions and cause dissipation.
We can also check this by evaluating the integrated occupations, i.e. by integrating over the occupation
distribution in the region $\omega\in[0,\Delta]$ and the region $\omega\in[\Delta,-\Delta+\Omega]$
\begin{eqnarray}
n_{E(in)}&=&\int_{0}^{\Delta}d\omega\left(n_{\uparrow}(0,\omega)+n_{\downarrow}(0,\omega)\right),\nonumber\\
n_{E(out)}&=&\int_{\Delta}^{-\Delta+\Omega}d\omega\left(n_{\uparrow}(0,\omega)+n_{\downarrow}(0,\omega)\right).
\label{eq:totExc_Pert}
\end{eqnarray}
We plot the integrated occupations for the equilibrium case and
for the driven non-equilibrium  steady case as a function of temperature in Fig. \ref{fig:totalOCC_Q0}. The case without periodic driving potential follows the
standard equilibrium statistical mechanics with vanishing total excitation at $T=0$. In the presence of driving potential, the total occupation at $T=0$ is
nonzero for both $n_{E(in)}$ and $n_{E(out)}$. With varying temperature, $n_{E(in)}$ only has small changes while $n_{E(out)}$ has larger change.
In addition, the occupation function shows a plateau between $\Delta$ and $-\Delta+\Omega$, and also shows small occupation in the region
$\omega\in (-\Delta+\Omega,\Delta+\Omega)$ with the similar structure to $\omega\in(-\Delta,\Delta)$.
In fact, the transitions induced by periodic driving interplay with the dissipation due to the superconducting bath (especially for energy above the
bulk superconducting gap); and this competition results in small finite occupation in the non-equilibrium stationary states.

\begin{figure}[t]
\centering
\includegraphics[width=3.4in,clip]{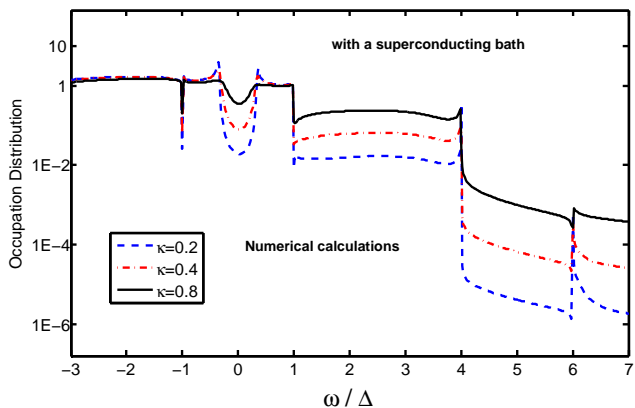}
\caption{The occupation distribution $n_{\uparrow}(0,\omega)+n_{\downarrow}(0,\omega)$ for the zeroth Floquet band from full
numerical solution of Eqs. (\ref{eq:MatrixEq_Self}-\ref{eq:Q0Matrix}). Following parameters were used: $\kappa=0.2,0.4,0.8$, $\Omega/\Delta=5.0$, $W/\Delta=0.4$,
temperature $T/\Delta=0.2$, and finite band width for the normal metal part $(-D,\, D)$ with $D/\Delta=3.5$. $N=30$ ($2N+1$ Floquet bands).}
\label{fig:OCCn0_Num}
\end{figure}

\begin{figure}[t]
\centering
\includegraphics[width=3.4in,clip]{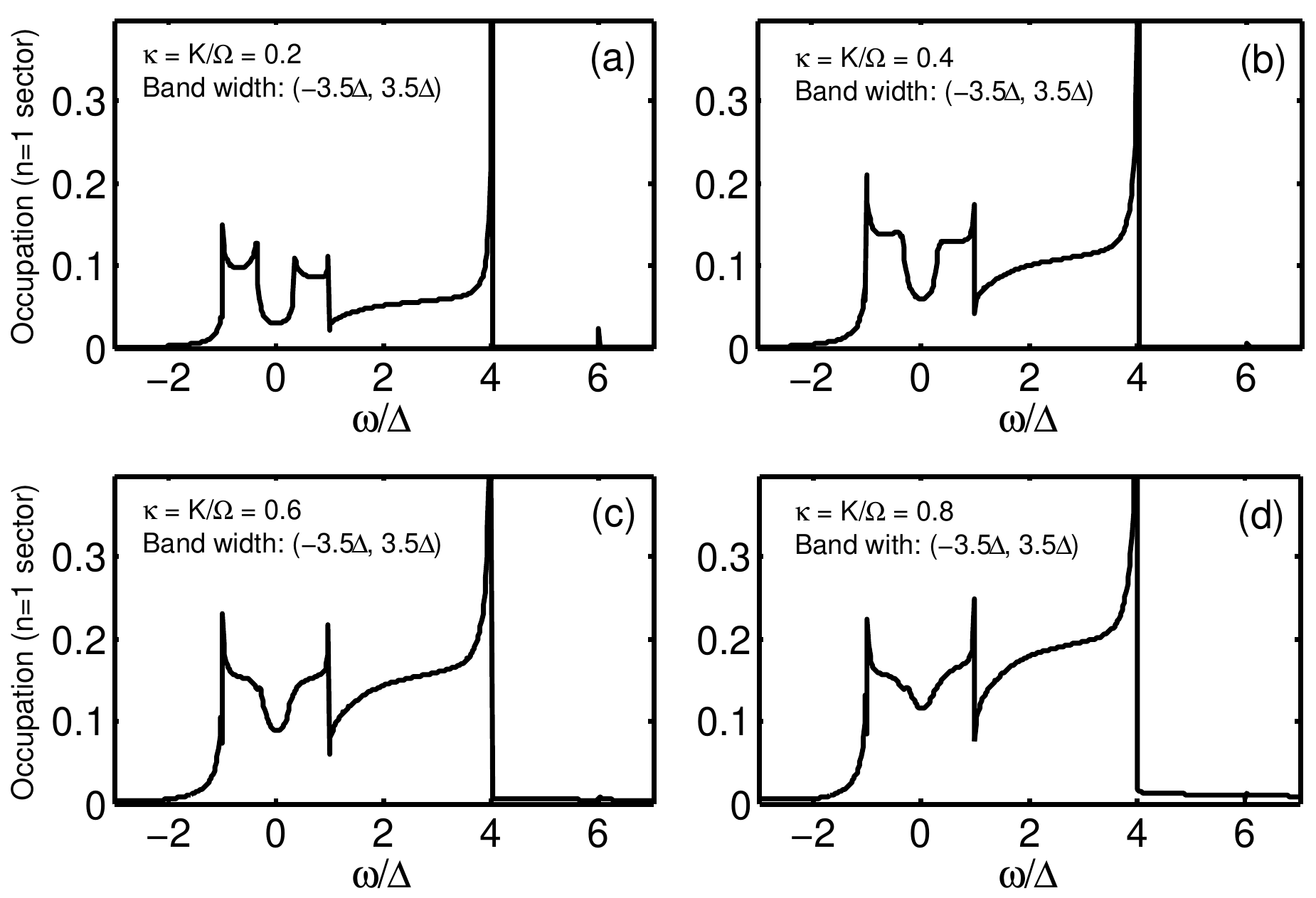}
\caption{The occupation distribution $|n_{\uparrow}(1,\omega)|+|n_{\downarrow}(1,\omega)|$ for the first Floquet band from the
numerical analysis of Eqs. (\ref{eq:MatrixEq_Self}-\ref{eq:Q0Matrix}). We choose $\Omega/\Delta=5.0$, $W/\Delta=0.4$, temperature $T/\Delta=0.2$,
and finite band width for the normal metal part $(-D,\, D)$ with $D/\Delta=3.5$. $N=30$ (totally $2N+1$ bands).}
\label{fig:OCCn1_Num}
\end{figure}

Finally, we consider the occupation distribution for the zeroth Floquet band from full solution of matrix Eqs. (\ref{eq:MatrixEq_Self}-\ref{eq:Q0Matrix}),
which is shown in Fig. \ref{fig:OCCn0_Num}. The occupation distribution below the positive bulk gap, i.e. $\omega<-\Delta+\Omega$,
is qualitatively similar to the perturbative results, namely there are large occupation for $\omega\in(\Delta_{\mathrm{ind}},\,\Delta)$,
and small finite occupation above the bulk gap $\Delta$. However, the gap structure appearing in the region $\omega\in(\Delta,\,-\Delta+\Omega)$
of the second order perturbation theory is smeared out due to the higher order processes. With increasing the driving amplitude,
the occupation in the regime $\omega>\Delta$ becomes larger and larger.  The numerical results for the fast oscillatory part from the first Floquet band $n(1,\omega)$ is shown in Fig. \ref{fig:OCCn1_Num}.

\section{Conclusions} \label{sec:conclusions}

In this paper, we developed non-equilibrium  Keldysh Green's function approach to study periodically modulated systems with dissipation.
As a practical example, we considered periodically driven normal system in contact with a superconducting bath.
After integrating out the fermionic bath degrees of freedom and incorporating their effects into  the self-energy, we can treat effect of dissipation as well as superconducting proximity
non-perturbatively. We obtained a set of kinetic equations, which have the same structural form as in equilibrium, with the important distinction that
the Green's functions and self-energy are now infinite-dimensional matrices in the Floquet space. By self-consistently truncating the matrices for a finite bandwidth or employing
perturbative treatment for the small driving amplitude, we computed various physical observables such as density of states and occupation distribution function for the zeroth Floquet band which can be measured in energy-resolved tunneling spectroscopy experiments. Perspectives for the transport measurements of the electron current shot noise
in periodically driven wires are also briefly discussed. Our new approach and results are useful for understanding the dissipative driven systems and non-equilibrium systems with proximity-induced superconductivity.

Our main findings for specific model considered here can be summarized as follows. In the case of a non-superconducting bath, we show that the steady-state distribution function is non-thermal with characteristic dependence on the driving frequency. In the case of a superconducting bath, only quasiparticles with energies above bulk gap can escape into the bath and thermalize. Below the bulk superconducting gap but above the induced gap, the steady-state occupation probability is enhanced because quasiparticles cannot escape into the superconducting bath. Thus, the steady-state distribution function exhibits complicated structure which depends on the induced and bulk superconducting gaps as well as driving frequency and amplitude. We also find that periodic driving modifies the proximity effect and leads to a finite density of states below the induced superconducting gap. This non-equilibrium effect becomes more significant with increasing the driving amplitude.

\section{Acknowledgements}

We are grateful to X. Li, M. F. Maghrebi, K. I. Seetharam, M. Vavilov for discussions. D.L. and R.L. acknowledge the hospitality of the Aspen Center for Physics supported by NSF grant No. PHY1066293,
where part of this work was done. This work at University of Wisconsin-Madison was financially supported in part by NSF Grants No. DMR-1606517, No. ECCS-1560732,
and by the Wisconsin Alumni Research Foundation. A.L. acknowledge hospitality of the Kavli Institute for Theoretical Physics, where parts of this work were completed and supported in part by the NSF under Grant No. NSF PHY11-25915.

\appendix

\section{Derivation of Green's functions $Q(n,\omega)$ and distribution function from perturbative expansion in $\kappa$}\label{app:Greenpert}
\begin{widetext}
In this Appendix, we show how to derive the Green's functions $Q(n,\omega)$ for $n=0, \pm 1$,
see Eqs. (\ref{eq:Q1pert})-(\ref{eq:Qkband0_2Opert}).
We can write Eq. (\ref{eq:FK_Q_self22}) in a matrix form:
\begin{equation}
\left(\begin{array}{ccccccc}
\ddots & \cdots & \cdots & 0 & 0 & 0 & 0\\
0 & -B_{1,2}(\omega) & \mathbb{I}-A_{1}(\omega) & -B_{1,0}(\omega) & 0 & 0 & 0\\
0 & 0 & -B_{0,1}(\omega) & \mathbb{I}-A_{0}(\omega) & -B_{0,-1}(\omega) & 0 & 0\\
0 & 0 & 0 & -B_{-1,0}(\omega) & \mathbb{I}-A_{-1}(\omega) & -B_{-1,-2}(\omega) & 0\\
0 & 0 & 0 & 0 & \cdots & \cdots & \ddots
\end{array}\right)\left(\begin{array}{c}
\vdots\\
\breve{Q}_{k}(2,\omega)\\
\breve{Q}_{k}(1,\omega)\\
\breve{Q}_{k}(0,\omega)\\
\breve{Q}_{k}(-1,\omega)\\
\breve{Q}_{k}(-2,\omega)\\
\vdots
\end{array}\right)=\left(\begin{array}{c}
\vdots\\
0\\
0\\
\breve{Q}_{0k}(\omega)\\
0\\
0\\
\vdots
\end{array}\right)
\end{equation}
where the functions are given by
\begin{eqnarray}
A_{n}(\omega) & = & \breve{Q}_{0k}(\omega+n\Omega)\breve{M}_{0}\breve{q}_{\rm{bath}}(\omega+n\Omega)\breve{M}_{0}^{*}\\
B_{n+1,n}(\omega) & = & \breve{Q}_{0k}(\omega+(n+1)\Omega)\breve{M}_{0}\breve{q}_{\rm{bath}}(\omega+(n+1)\Omega)\breve{M}_{-1}^{*}+\breve{Q}_{0k}(\omega+(n+1)\Omega)\breve{M}_{1}\breve{q}_{\rm{bath}}(\omega+n\Omega)\breve{M}_{0}^{*}\\
B_{n-1,n}(\omega) & = & \breve{Q}_{0k}(\omega+(n-1)\Omega)\breve{M}_{0}\breve{q}_{\rm{bath}}(\omega+(n-1)\Omega)\breve{M}_{1}^{*}+\breve{Q}_{0k}(\omega+(n-1)\Omega)\breve{M}_{-1}\breve{q}_{\rm{bath}}(\omega+n\Omega)\breve{M}_{0}^{*}.
\end{eqnarray}
Notice that $B_{n\pm1,n}(\omega)\sim\kappa$. Thus, one can simplify above equations by performing perturbative expansion in $\kappa\ll 1$.
Using the matrix identity $(A-\kappa B)^{-1}=A^{-1}+\kappa A^{-1}BA^{-1}+O(\kappa^{2})$,
we obtain the Green's function to the leading order of $\kappa$, see Eqs. (\ref{eq:Q1pert})-(\ref{eq:Qkband0_2Opert}).
\end{widetext}

Next, we simplify the distribution matrix by noticing that
\begin{equation}
F(t,\omega)=F_{0}(\omega)+\kappa F_{1}(t,\omega)+O(\kappa^{2})
\end{equation}
where after Fourier expansion
\begin{eqnarray}
F(0,\omega) & = & F_{0}(\omega)+\kappa F_{1}(0,\omega)+O(\kappa^{2}),\\
F(n\neq0,\omega) & = & \kappa F_{1}(n,\omega)+O(\kappa^{2}).
\end{eqnarray}
Therefore, one can show that $\breve{Q}_{k}(\pm1,\omega)\sim \kappa$ whereas $\breve{Q}_{k}(0,\omega)$ (equilibrium Green's function) is order one. Having this in mind, one can now expand Eq. (17) to find
\begin{eqnarray}
\breve{Q}_{k}^{K}(0,\omega) & = & \breve{Q}_{k}^{R}(0,\omega)F_{0}(\omega)-F_{0}(\omega)\breve{Q}_{k}^{A}(0,\omega)\\
F_{1}(0,\omega) & = & 0\\
\breve{Q}_{k}^{K}(1,\omega) & = & \breve{Q}_{k}^{R}(0,\omega+\Omega)\kappa F_{1}(1,\omega)\nonumber\\
       && -F_{0}(\omega+\Omega)\breve{Q}_{k}^{A}(1,\omega)+\breve{Q}_{k}^{R}(1,\omega)F_{0}(\omega)\nonumber\\
       &&-\kappa F_{1}(1,\omega)\breve{Q}_{k}^{A}(0,\omega)\\
\breve{Q}_{k}^{K}(-1,\omega) & = & \breve{Q}_{k}^{R}(0,\omega-\Omega)\kappa F_{1}(-1,\omega)\nonumber\\
    &&-F_{0}(\omega-\Omega)\breve{Q}_{k}^{A}(-1,\omega)+\breve{Q}_{k}^{R}(-1,\omega)F_{0}(\omega)\nonumber\\
    && -\kappa F_{1}(-1,\omega)\breve{Q}_{k}^{A}(0,\omega).
\end{eqnarray}
One may notice that $F_{0}(\omega)$ is, in fact, the equilibrium distribution
function. The functions $F_{1}(\pm1,\omega)$ can be obtained recursively. Note that
due to the relation $\underline{F}^{\dagger}=\underline{F}$, one can show that
$F(1,\omega)^{*}=F(-1,\omega+\Omega)$. 

\begin{widetext}
\section{Proximity effect and DOS in the case of a superconducting bath}\label{app:AnalyticExpression}

In this appendix, we provide details of the derivation of the DOS for energies below the induced gap. Specifically, we derive the functions $\nu^{(2)}(0,\epsilon=0)$ and $\nu^{(2)}(1,\epsilon=0)$ defined in Eqs. (\ref{eq:nuN0}) and (\ref{eq:nuN1}), respectively.

Let us first consider $\nu^{(2)}(0,\omega=0)$, defined in Eq. (\ref{eq:nuN0}).  Analytical results can be obtained in two different regimes: $\Omega \ll \Delta$ and $\Omega \gg \Delta$ respectively.
Note that in order to obtain the correct analytical results, we have to carefully take the $\eta\rightarrow 0^+$ limit at the end of the calculation. For $\Omega\ll \Delta$ case, one finds
\begin{equation}
 \nu^{(2)}(0,\omega=0) = \Re \left[ \frac{2 \Gamma ^2 \Delta  \left(\Omega ^2 \left(-12 \Gamma ^2-\Delta ^2+\Omega ^2\right)+\left(\sqrt{\Delta ^2-\Omega ^2}+\Delta \right)
   \left(8 \Gamma ^3+12 \Gamma ^2 \Delta +6 \Gamma  \left(\Delta ^2-\Omega ^2\right)+\Delta ^3-\Delta  \Omega ^2\right)\right)}{\Omega
   \sqrt[4]{\Delta ^2-\Omega ^2} \left(2 \Gamma +\sqrt{\Delta ^2-\Omega ^2}\right)^4 \sqrt{\Gamma ^2 \left(-\sqrt{\Delta ^2-\Omega
   ^2}\right)+2 \Gamma  \Omega ^2+\Omega ^2 \sqrt{\Delta ^2-\Omega ^2}}}\right],
\end{equation}
where $\Gamma=\pi\rho_F W^2$. In the opposite limit $\Omega\gg \Delta$, we obtain
\begin{equation}
  \nu^{(2)}(0,\omega=0) =\Re[ D_{L1}+D_{L2} ],
\end{equation}
where
\begin{eqnarray}
 D_{L1} &=& \Bigg\{ 2\Gamma^2 \Big( \Gamma ^2 (\Delta -\Omega ) (\Delta +\Omega ) \left(\Delta  \left(\sqrt{(\Delta -\Omega ) (\Delta +\Omega )}+\Delta \right)-\Omega ^2\right)\nonumber\\
 &&+ \Omega  \left((2 \Gamma +\Delta ) \left(\sqrt{\Delta ^2-\Omega ^2}+\Delta \right)-\Omega ^2\right) \left(\Delta  \sqrt[4]{\Delta ^2-\Omega ^2}
   \sqrt{\left(\Omega ^2-\Gamma ^2\right) \sqrt{\Delta ^2-\Omega ^2}+2 \Gamma  \Omega ^2}-\Delta ^2 \Omega +\Omega ^3\right) \Big)\Bigg\}\nonumber\\
 && \Bigg/ \Bigg\{ \Omega ^2 \sqrt{(\Delta -\Omega ) (\Delta +\Omega )} \left(2 \Gamma +\sqrt{(\Delta -\Omega ) (\Delta +\Omega )}\right)^2 \nonumber\\
 && \quad\quad \times \left(\Gamma ^2
   \left(-\sqrt{(\Delta -\Omega ) (\Delta +\Omega )}\right)+2 \Gamma  \Omega ^2+\Omega ^2 \sqrt{(\Delta -\Omega ) (\Delta +\Omega )}\right) \Bigg\},
\end{eqnarray}
and
\begin{equation}
  D_{L2}=\frac{\frac{2 \Gamma ^2 \Delta ^2}{\Omega ^2}+\frac{\Delta  \Omega  (2 \Gamma +\Delta ) \sqrt{\Omega ^2-\Delta ^2}}{4 \Gamma ^2-\Delta
   ^2+\Omega ^2}-\frac{32 \Gamma ^5 (2 \Gamma +\Delta )}{\left(4 \Gamma ^2-\Delta ^2+\Omega ^2\right)^2}+\frac{8 \Gamma ^3 (\Gamma +\Delta )}{4
   \Gamma ^2-\Delta ^2+\Omega ^2}+\frac{2 \Gamma  (2 \Gamma +\Delta ) \sqrt{\Omega ^2-\Delta ^2}}{\Omega }}{(2 \Gamma +\Delta )^2}.
\end{equation}

Similarly, one can calculate analytical expression for $\nu^{(2)}(1,\omega=0)$, defined in Eq. (\ref{eq:nuN1}):
\begin{equation}
 \nu^{(2)}(1,\omega=0) = \text{Re}\left[\nu_{1,\text{NUM}} /\nu_{1,\text{DEN}}\right],
\end{equation}
where
\begin{eqnarray}
 \nu_{1,\text{NUM}}&=&-\Gamma  \Bigg[ 4 \Gamma ^4 \left(2 \Delta ^2-\Omega ^2\right) \left(\Delta ^3+\Delta ^2 \sqrt{\Delta ^2-\Omega ^2}+\Omega ^2 \sqrt{\Delta ^2-\Omega
   ^2}-\Delta  \Omega ^2\right)+\Omega ^2 \left(\Delta ^2-\Omega ^2\right)^{7/2} \nonumber\\
   &&+4 \Gamma ^3 \left(\Delta ^2-\Omega ^2\right) \left(3 \Delta ^4+3 \Delta ^2 \Omega ^2-2 \Delta  \Omega ^2 \sqrt{\Delta ^2-\Omega ^2}+3 \Delta
   ^3 \sqrt{\Delta ^2-\Omega ^2}-3 \Omega ^4\right)\nonumber\\
   &&+\Gamma  \left(\Delta ^2-\Omega ^2\right)^2 \left(\Delta ^4+6 \Delta ^2 \Omega ^2-\Delta  \Omega ^2 \sqrt{\Delta ^2-\Omega ^2}+\Delta ^3
   \sqrt{\Delta ^2-\Omega ^2}-6 \Omega ^4\right)\nonumber\\
   &&+\Gamma ^2 \left(\Delta ^2-\Omega ^2\right) \left(6 \Delta ^5-11 \Delta ^3 \Omega ^2+13 \Delta ^2 \Omega ^2 \sqrt{\Delta ^2-\Omega ^2}-13
   \Omega ^4 \sqrt{\Delta ^2-\Omega ^2}+6 \Delta ^4 \sqrt{\Delta ^2-\Omega ^2}+5 \Delta  \Omega ^4\right)\Bigg],\\
 \nu_{1,\text{DEN}}&=& \Omega  \sqrt{\Delta ^2-\Omega ^2} \left(4 \Gamma ^2+4 \Gamma  \sqrt{\Delta ^2-\Omega ^2}+\Delta ^2-\Omega ^2\right) \sqrt{\Gamma ^2
   \left(\Omega ^2-\Delta ^2\right)+2 \Gamma  \Omega ^2 \sqrt{\Delta ^2-\Omega ^2}+\Omega ^2 \left(\Delta ^2-\Omega ^2\right)}\nonumber\\
   &&\times \left[\Gamma ^2 \left(2 \Omega ^2-4 \Delta ^2\right)+\Gamma  \sqrt{\Delta ^2-\Omega ^2} \left(3 \Omega ^2-4 \Delta ^2\right)-\left(\Delta ^2-\Omega
   ^2\right)^2\right].
\end{eqnarray}

\end{widetext}

\bibliography{FloquetK1,FloquetETH_OPEN,topological_wires11}

\end{document}